%% file: jrnl_main.tex
\PassOptionsToPackage{hidelinks}{hyperref}
\documentclass[lettersize,journal]{IEEEtran}
\usepackage{amsmath,amsfonts}
\usepackage{algorithm}
\usepackage{array}
\usepackage[caption=false,font=footnotesize,labelfont=sf,textfont=sf]{subfig}
\usepackage{textcomp}
\usepackage{stfloats}
\usepackage{url}
\usepackage{verbatim}
\usepackage{graphicx}
\usepackage{cite}
\usepackage{multirow}
\usepackage{siunitx}
\usepackage{adjustbox}
\usepackage{amssymb}

\usepackage{comment}

\usepackage{algpseudocode}

\usepackage[nolist]{acronym}

\usepackage{tikz}
\usepackage{pgfplots}
\usepgfplotslibrary{fillbetween}
\usetikzlibrary{calc}
\pgfplotsset{compat=1.18}

\usepackage{orcidlink}

\usetikzlibrary{positioning, arrows.meta, shapes}

\begin{document}

\title{Convergence Time Distributions for Max-Consensus over Unreliable Networks}

\author{Katharina~Stich~\orcidlink{0009-0003-6643-9462}~\IEEEmembership{Graduate Student Member, IEEE}, Bastian~Perner~\orcidlink{0009-0006-1285-2229}~\IEEEmembership{Graduate Student Member, IEEE}, Friedemann~Laue~\orcidlink{0000-0002-7477-8312}~\IEEEmembership{Graduate Student Member, IEEE}, Torsten~Reissland~\orcidlink{0000-0002-2263-3875}~\IEEEmembership{Member, IEEE}, and Norman~Franchi~\orcidlink{0000-0002-2777-4722}~\IEEEmembership{Member, IEEE}
\thanks{This work has been supported by the Federal Ministry of Research, Technology and Space of the Federal Republic of Germany as part of the project Open6GHub+ under contract 16KIS2404, and 6G-Sensoria.}%
\thanks{The authors are with the Institute of Smart Electronics and Systems, Friedrich-Alexander Universit{\"a}t Erlangen-N{\"u}rnberg, 91054 Erlangen, Germany (e-mail: \{katharina.stich, bastian.perner, friedemann.laue, torsten.reissland, norman.franchi\}@fau.de).}%
\thanks{This work has been submitted to the IEEE for possible publication. Copyright may be transferred without notice, after which this version may no longer be accessible.}}


\markboth{Journal,~Vol.~XX, No.~YY, Month~Year}%
{Shell \MakeLowercase{\textit{et al.}}: A Sample Article Using IEEEtran.cls for IEEE Journals}


\maketitle

\begin{acronym}
    \acro{MAS}{multi-agent system}
    \acro{MASs}{multi-agent systems}
    \acro{UAV}{unmanned aerial vehicle}
    \acro{PMF}{probability mass function}
    \acro{CDF}{cummulative distribution function}
    \acro{EV}{expected value}
    \acro{LiFE-CD}{Link Failure Effects on Consensus Distributions}
\end{acronym}

\begin{abstract}
This paper proposes the \acf{LiFE-CD} algorithm for convergence time analysis of the max-consensus algorithm in multi-agent systems under Bernoulli-distributed link failures. Unlike existing approaches, which either assume ideal communication or provide asymptotic upper bounds on the expected convergence time, \ac{LiFE-CD} deterministically computes the full probability distribution of the convergence time from network topology and individual link failure probabilities, without simulation. The full probability distribution enables deadline-aware protocol design with specified reliability guarantees. Based on geometrically distributed link delays, the proposed algorithm iteratively reduces the given network topology considering both unicast and broadcast transmissions. \ac{LiFE-CD} yields exact results for acyclic networks and, for cyclic networks, tight upper bounds on the convergence time via shortest-path spanning tree construction. Numerical results confirm analytical exactness for acyclic networks, validate tightness for cyclic networks, and demonstrate improvement over existing approaches. Our complexity analysis shows reduced computational cost compared to Monte Carlo simulations, while eliminating stochastic variability and enhancing reproducibility. All results extend directly to min-consensus by structural equivalence.
\end{abstract}


\section{Introduction}
\label{sec:introduction}
\input{introduction_new}

\section{Preliminaries}
This section introduces the notation used in this work and provides the necessary background in graph theory, probability theory, and geometric 
distributions.
\label{sec:preliminaries}
\input{notation}

\section{Max-Consensus}
\label{sec:consensus}
This section reviews the standard max-consensus protocol and its convergence properties. We first present the protocol specification for both ideal and unreliable communication links \cite{Katha, Molinari2, NoisyLinks}, followed by a summary of existing convergence time analyses.
\subsection{Protocol}
\subsubsection{Ideal Communication Links}
\input{standard}

\subsubsection{Unreliable Communication Links}
\input{consensus}
\label{sec:consensus_unreliable}
\subsection{Convergence}
\input{convergence}

\section{Proposed Approach}
\label{sec:approach}
This section formally describes the problem at hand and derives the delay model, 
considering unicast and broadcast transmissions, non-geometric 
delay compositions, and the influence of network topology. Moreover, 
the proposed \ac{LiFE-CD} algorithm is presented and its working principle 
is illustrated through a representative example.
\input{problem_new}
\input{system_new}
\subsection{Illustrative Example}
\label{ssec:example}
\input{example}

\section{Performance Evaluation}
This section evaluates \ac{LiFE-CD} in terms of convergence time distributions for acyclic and cyclic networks, expected convergence time compared to state-of-the-art bounds, and computational complexity relative to Monte Carlo simulation. All Monte Carlo results are averaged over $5{,}000$ independent runs.
\label{sec:evaluation}
\input{eval2}

\input{eval1}

\subsection{Computational Complexity Analysis}
\input{performance_2}

\section{Conclusion}
\label{sec:conclusion}
\input{conclusion}

\section*{Acknowledgment}
The work contributes to the research within the 6G-Valley innovation cluster. The authors used the AI language tool Claude (Anthropic) for grammar and spelling checks and to improve the readability of the manuscript. The authors take full responsibility for the content of this work.

\bibliographystyle{IEEEtran}
\bibliography{journal3}

\begin{IEEEbiography}[{\includegraphics[width=1in,height=1.25in,clip,keepaspectratio]{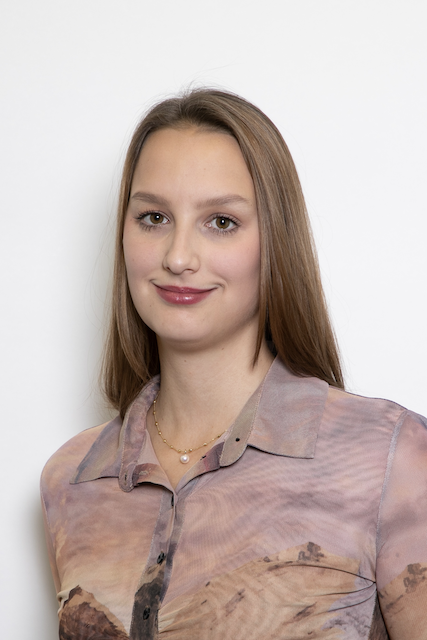}}]{Katharina Stich} (Graduate Student Member, IEEE) received the B.Sc. and M.Sc. degrees in electrical engineering with a specialization in information technology from Friedrich-Alexander-Universit\"{a}t Erlangen-N\"{u}rnberg (FAU), Erlangen, Germany, in 2021 and 2023, respectively. In November 2023, she joined the Institute for Smart Electronics and Systems (LITES) at FAU, where she is currently pursuing the Ph.D. degree. Her research interests include distributed networked control systems, decision-making in multi-agent systems with focus on consensus algorithms, and 6G communication networks.
\end{IEEEbiography}

\begin{IEEEbiography}[{\includegraphics[width=1in,height=1.25in,clip,keepaspectratio]{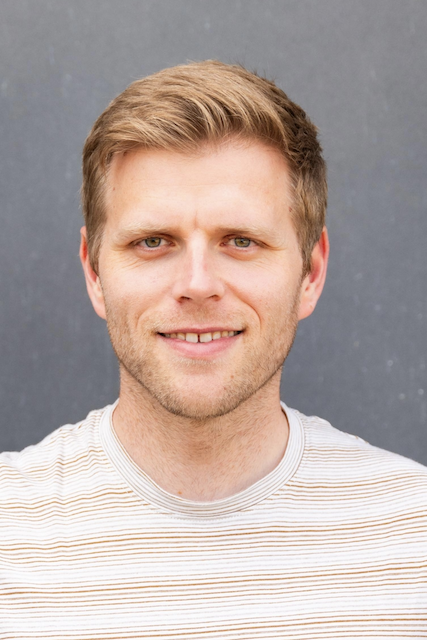}}]{Bastian Perner}(Graduate Student Member, IEEE) received the B.Sc. and M.Sc. degrees in Industrial Mathematics with specialization in optimzation from Friedrich-Alexander-Universit\"{a}t Erlangen-N\"{u}rnberg (FAU), Erlangen, Germany, in 2013 and 2016, respectively. In 2017, he joined the Fraunhofer Institute of Integrated Circuits (IIS) as a Research Assistant participating in various research projects and 3GPP standardization. In 2023, he joined the Chair of Electrical Smart City Systems, now part of the Institute for Smart Electronics and Systems, FAU, as a Research Assistant in the Radio Technology and Networks Group. His research interests include localization, channel estimation, and distributed radio sensing.
\end{IEEEbiography}

\begin{IEEEbiography}[{\includegraphics[width=1in,height=1.25in,clip,keepaspectratio]{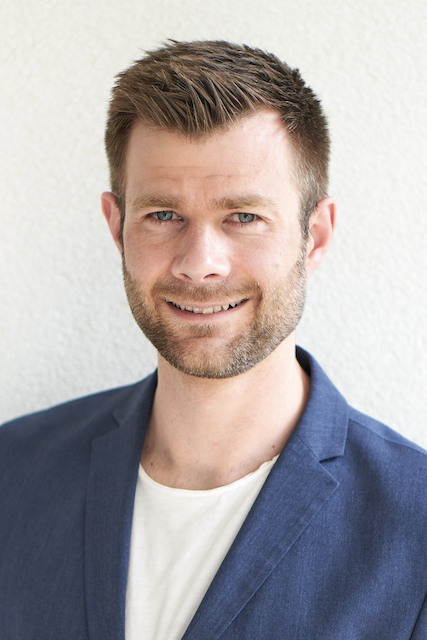}}]{Friedemann Laue} (Graduate Student Member, IEEE) received the B.Sc. and M.Sc. degrees (Hons.) in computer and communications systems engineering from Technische Universität Braunschweig, Germany, in 2011 and 2015, respectively. From 2015 to 2017, he started his professional career as a Signal Processing Engineer with Capical GmbH, Braunschweig. From 2017 to 2024, he worked at the Fraunhofer Institute for Integrated Circuits (IIS) and the Institute for Digital Communications at Friedrich-Alexander-Universit\"{a}t Erlangen-N\"{u}rnberg (FAU), Germany, participating in various research projects on mobile communication systems, including his PhD research on Reconfigurable Intelligent Surfaces. In 2024, he joined the Institute for Smart Electronics and Systems, FAU, where his current research focuses on resilient communication systems.
\end{IEEEbiography}

\begin{IEEEbiography}[{\includegraphics[width=1in,height=1.25in,clip,keepaspectratio]{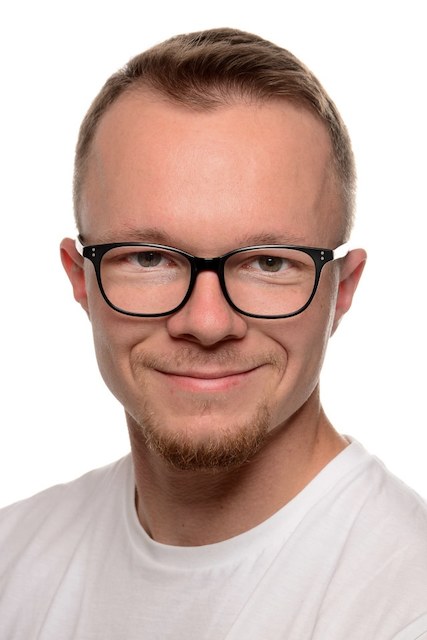}}]{Torsten Reissland}(Member, IEEE) completed his Bachelor’s degree in computer science and systems engineering at Ilmenau Technical University in 2013. He successfully completed his Master’s degree in electrical engineering at Friedrich-Alexander-Universit\"{a}t
Erlangen-N\"{u}rnberg (FAU) in March 2016. From May 2016 to July 2023, he worked as a research assistant at the Institute for Electronics Engineering (LTE) at FAU. From May 2022 to July 2023, he also headed the Circuits, Systems and Hardware Test group at this institute. In July 2023, he completed his doctorate with distinction. He holds his current position as the leader of the Algorithms and Digital Signal Processing group at the Institute for Smart Electronics and Systems at FAU since November 2023. Dr. Reissland has been engaged in research and development of radar systems, JCAS systems and 6G communication systems. 
\end{IEEEbiography}

\begin{IEEEbiography}[{\includegraphics[width=1in,height=1.25in,clip,keepaspectratio]{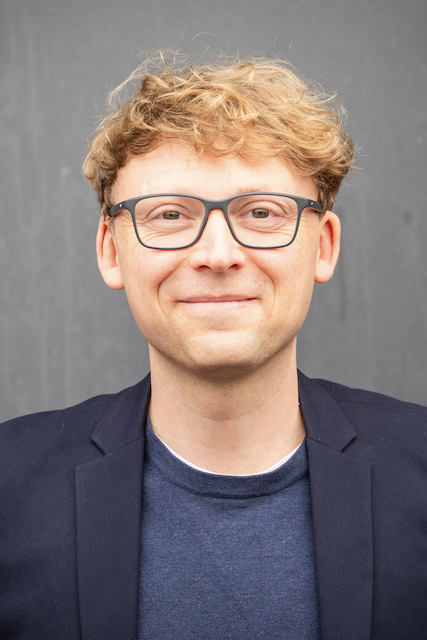}}]{Norman Franchi} (Member, IEEE) received the Dipl.-Ing. (M.S.)
and Dr.-Ing. (Ph.D.) degrees in electrical engineering from Friedrich-
Alexander-Universit\"{a}t Erlangen-N\"{u}rnberg (FAU), Germany, in 2007 and
2015, respectively. He is currently a Full Professor (W3) and the
Head of the Institute for Smart Electronics and Systems (LITES), FAU.
From 2007 to 2011, he was with the Automotive Research and Development
Sector as a System and Application Engineer for advanced networked
control system design. From 2012 to 2015, he was a Research Associate
with the Institute for Electronics Engineering, FAU, focused on SDR-
based V2X communications. From 2015 to 2021, he was with Dresden
University of Technology as a Senior Research Group Lead for resilient
mobile communications systems and 5G industrial networks. His research
interests include 6G, AI-native 6G, ISAC, mixed-signal, RFIC, and systems
design (Sub 6 GHz to THz), and resilient and trustworthy communications
circuits and systems. He is a member of the IEEE ISAC Initiative, Open 6G
Hub Germany, and 6G Platform Germany. He is a Scientific Advisory Board
Member of KI Park e.V., Germany.
\end{IEEEbiography}
\end{document}

%% file: introduction_new.tex
\IEEEPARstart{C}{onsensus} protocols are fundamental to decentralized coordination in \ac{MASs}, where autonomous agents interact and collaborate to agree on joint decisions through iterative local communication~\cite{Dimakis2010,information}. With the rapid deployment of autonomous systems, from multi-robot warehouses to \ac{UAV} swarms and vehicle platoons, reliable distributed coordination under uncertain communication conditions has become increasingly critical~\cite{TSIPN5, SP1}. In consensus protocols, agents share their current information state with their neighbors and update their value based on the information received until all agents converge to a common value. One of the consensus algorithms most studied is average consensus, where agents iteratively converge to the average of their initial values, with applications in formation control, flocking, and rendezvous timing~\cite{Olfati,AV1}.

In contrast, many applications require max-/min-consensus protocols, which enable agents to converge to the maximum or minimum of the initial information states across all agents. These applications include target interception~\cite{AP2} and shortest-path planing\cite{AP3} in \ac{UAV} swarms, distributed target tracking~\cite{TargetTracking,Farahmand2011}, energy management in smart grids~\cite{energy1,AP7,AP10}, data aggregation in wireless sensor networks~\cite{noiseConsensus1, randomNetwork1}, and distributed Kalman filtering~\cite{Kalman,AP8,AP9}.

The performance of max-/min-consensus algorithms depends on frequent and reliable communication. While existing research addresses various communication constraints, including wireless signal superposition~\cite{Molinari1,Molinari2,Molinari3}, quantization and limited bandwidth~\cite{Quantize}, asynchrony and communication delays~\cite{asynchronousConsensus,asynchronousConsensus2,asynchronousConsensus3}, and channel noise~\cite{noiseConsensus1,noiseConsensus2,noiseConsensus3,NoisyLinks}, packet losses and stochastic link failures remain a fundamental challenge. In dynamic environments with unstable wireless links, temporary disconnections can significantly delay or disrupt the consensus process. For time-critical applications such as collision avoidance in autonomous vehicles or real-time coordination in industrial robotics, predicting the expected convergence time under probabilistic failures is essential for safe and reliable system operation.

\subsection{Related Work}
Early works on max-/min-consensus established convergence conditions for deterministic networks. Using max-plus algebra, the authors of \cite{MaxCSS} showed that max-consensus convergence is bounded by the diameter of the graph for strongly connected time-invariant networks, with applications in leader election and rendezvous. This was extended by the work in \cite{Nejad2} to time-varying switching topologies, proving that a jointly strongly connected network is sufficient for convergence.

In addition to the analysis of deterministic communication links, the effect of stochastic link failures has been characterized by probabilistic convergence guarantees. For example in \cite{randomNetwork1}, the convergence of max-consensus was proven for directed networks with random link failures in four probabilistic senses (almost surely, in probability, in expectation, and mean square), which are equivalent when the expected graph is strongly connected. Related consensus frameworks under probabilistic failures provide similar convergence guarantees, including almost-sure and mean-square convergence, for unified min-max approaches \cite{randomNetwork+} and group consensus protocols \cite{groupConsensus}. However, these analyses focuses on convergence existence rather than on the time required to reach consensus.

Alternative approaches achieve robustness against failures through modified variants of the standard max-/min-consensus protocol. The algorithm proposed in~\cite{MaxCUCL} employs acknowledgment-based feedback channels to guarantee finite-time convergence despite packet drops. In \cite{wirelessConsensus} and \cite{Broadcast2}, random-pairwise-max and random-broadcast-max protocols are proposed, where in each iteration, a single node activates randomly and shares its information either with one neighbor or with all neighbors. Based on these protocols, asymptotic bounds on the expected convergence time under probabilistic failures have been derived. However, they fundamentally differ from the standard max-/min-consensus protocol through sequential rather than simultaneous node activation.

Most closely related to our work are studies analyzing the expected convergence time of the standard max-/min-consensus protocol under probabilistic link failures. The authors of \cite{Katha} investigate the impact of packet errors on min-consensus convergence time in industrial scenarios through simulation, but do not provide general analytical models. Of particular relevance is the study of \cite{bernoulli}, which analyzed the standard max-consensus protocol in time-varying networks with Bernoulli link dropouts. The authors proved finite-time convergence with probability one and derived asymptotic upper bounds on the expected convergence time and its dispersion. Nevertheless, the results remain limited to asymptotic bounds and do not provide an exact, topology-dependent computation of the convergence time. Furthermore, no probability distribution of the convergence time is provided, which would enable predictions on the likelihood of reaching consensus within a given deadline, which is critical for time-sensitive applications.

\subsection{Contributions and Paper Organization}
Despite significant progress, no existing work provides an exact, topology-dependent computation of the consensus time for the standard max-/min-consensus protocol under probabilistic link failures. Specifically, to the best of the authors knowledge, an algorithm for computing the probability distribution of the consensus convergence time does not yet exist in the literature. To provide reliable performance predictions, such an algorithm must account for the underlying network topology, which is particularly important for time-critical application.

This paper addresses this gap by proposing the \acf{LiFE-CD} algorithm. By modeling communication delays through geometric distributions and explicitly distinguishing between unicast and broadcast transmissions, \ac{LiFE-CD} integrates stochastic analysis with graph-theoretic reasoning. The key contributions of the paper can be summarized as follows:
\begin{itemize}
    \item We propose a deterministic computation algorithm that recursively combines geometric delay distributions through convolution operations for unicast transmissions and maximum operations for broadcast transmissions.
    \item We derive exact convergence times for acyclic networks and tight upper bounds for cyclic networks via shortest-path spanning tree construction.
    \item We provide full probability distribution of convergence time, enabling predictions on the likelihood of consensus within a given deadline—critical for time-critical applications.
    \item We investigate computational efficiency and demonstrate improvements over Monte Carlo simulation-based approaches.
\end{itemize}

These capabilities make \ac{LiFE-CD} particularly well-suited for time-critical applications in industrial robotics, autonomous vehicles, \ac{UAV} swarms, and sensor networks, where reliable and predictable consensus timing is essential for system-wide coordination and safety. Due to structural equivalence, the proposed method is applicable to both max- and min-consensus protocols.

The remainder of this paper is organized as follows.
Section~\ref{sec:preliminaries} establishes the graph-theoretic and probabilistic foundations upon which the analysis builds. Section~\ref{sec:consensus} presents the max-consensus protocol and its convergence properties under both ideal and probabilistic communication links. Building on this, Section~\ref{sec:approach} formalizes the problem, introduces the proposed \ac{LiFE-CD} algorithm, and illustrates its operation through a representative example.
The algorithm is evaluated in Section~\ref{sec:evaluation}, where we discuss the exact \ac{PMF} and \ac{CDF} of the consensus convergence time for exemplary acyclic and cyclic topologies, demonstrating the advantages of the proposed \ac{LiFE-CD} algorithm compared to existing methods. Finally, Section~\ref{sec:conclusion} concludes the paper and outlines future research directions.

%% file: notation.tex
\subsection{Notation and Graph Theory}

\subsubsection{Basic Notation}
The set of real numbers is denoted by $\mathbb{R}$, the set of nonnegative real numbers by $\mathbb{R}_{0}$, the set of positive integers by $\mathbb{N}$, and the set of nonnegative integers by $\mathbb{N}_0$. The cardinality of a set $\mathcal{S}$ is denoted by $|\mathcal{S}|$. The discrete convolution operator is denoted by $\ast$, and the pointwise maximum operator by $\max(\cdot)$. The probability of an event is denoted by $\Pr(\cdot)$.

\subsubsection{Graph Definitions}
A graph is defined as $\mathcal{G} = (\mathcal{N}, \mathcal{E})$, where $\mathcal{N} = \{1, \ldots, n\} \subseteq \mathbb{N}$ is the node set with $n = |\mathcal{N}|$, and $\mathcal{E} \subseteq \mathcal{N} \times \mathcal{N}$ is the edge set. An edge $e_{ij} = (i,j) \in \mathcal{E}$ represents a directed communication link from node $i$ to node $j$. The neighbor set of node $i$ is $\mathcal{N}_i = \{j \in \mathcal{N} \mid (i,j) \in \mathcal{E}\}$.

A graph is undirected if $(i,j) \in \mathcal{E}$ implies $(j,i) \in \mathcal{E}$ for all $i,j \in \mathcal{N}$; otherwise, it is a directed graph. Throughout this work, we assume an undirected graph. A weighted graph is denoted by $\mathcal{G} = (\mathcal{N}, \mathcal{E}, W)$, where $W = [w_{ij}]$ is the weighted adjacency matrix with edge weights $w_{ij} \in \mathbb{R}_{0}$. For an undirected graph, W is symmetric, i.e., $w_{ij}=w_{ji}$ for all $i,j \in \mathcal{N}$. The out-degree of node $i$ is denoted defined as $\deg(i)= |\{j \in \mathcal{N} | (i,j) \in \mathcal{E}\}|$.

A path from node $v_1 \in \mathcal{N}$ to node $v_n \in \mathcal{N}$ is a sequence of nodes $P = (v_1, v_2, \ldots, v_n)$ with $n \geq 2$ such that $(v_k, v_{k+1}) \in \mathcal{E}$ for all $k = 1, \ldots, n-1$. We denote by $\mathcal{E}(P)$ the set of edges comprising path $P$, and write $|\mathcal{E}(P)| = n-1$ for the path length.

Let $d(i,j)$ denote the distance between nodes $i$ and $j$, defined as the length of the shortest path from $i$ to $j$. The eccentricity of node $i$ is defined by $e(i) = \max_{j \in \mathcal{N}} d(i,j)$, representing the maximum distance from $i$ to any other node. The diameter of the graph is given by $D = \max_{i \in \mathcal{N}} e(i)$, characterizing the maximum distance between any pair of nodes.

A directed graph is strongly connected, and an undirected graph connected, if a path exists between every pair of nodes. Throughout this work, we assume a connected graph. A cycle is a path $(v_1, v_2, \ldots, v_n)$ with $n \geq 3$ such that $v_1 = v_n$ and $v_1, \ldots, v_{n-1}$ are distinct. A sequence $(v_r, \ldots, v_s)$ is a subpath of $(v_1, \ldots, v_n)$, denoted $(v_r, \ldots, v_s) \sqsubseteq (v_1, \ldots, v_n)$, if $1 \leq r < s \leq n$ and $(v_k, v_{k+1}) \in \mathcal{E}$ for all $k = r, \ldots, s-1$.

\subsection{Probability Theory}
\label{sec:probability}
\subsubsection{Discrete Random Variables}
For a discrete random variable $X$ taking values in $\mathbb{N}_0$, let $f_X(k) = \Pr(X = k)$ denote the \acf{PMF} and $F_X(k) = \Pr(X \le k)$ the \acf{CDF} of $X$. These satisfy
\begin{align}
    f_X(k) &= \begin{cases}
        F_X(0), & k = 0, \\
        F_X(k) - F_X(k-1), & k \ge 1,
    \end{cases}
    \label{eq:PMF} \\
    F_X(k) &= \sum_{i=0}^k f_X(i),
    \label{eq:CDF}
\end{align}
and the expected value is given by
\begin{align}
    \mathbb{E}[X] = \sum_{k=0}^\infty k \, f_X(k).
    \label{eq:EV}
\end{align}

\subsubsection{Transmission Delay and Geometric Distribution}
\label{sec:GEO}
The transmission delay over edge $e_{ij}$ is modeled by a discrete random variable $X_{ij}$. With link failure probability $p_{ij}\in [0,1)$, $X_{ij}$ follows a geometric distribution representing the number of independent Bernoulli trials until the first transmission success. The  \ac{PMF}, \ac{CDF}, and expected value are respectively given by

\begin{align}
    f_{X_{ij}}(k) &= p_{ij}^{k-1}(1-p_{ij}), \quad k \geq 1,
    \label{eq:PMF_geometric} \\
    F_{X_{ij}}(k) &= 1 - p_{ij}^{k}, \quad k \geq 0,
    \label{eq:CDF_geometric} \\
    \mathbb{E}[X_{ij}] &= \frac{1}{1-p_{ij}}.
    \label{eq:EV_geometric}
\end{align}

%% file: standard.tex
The standard max-consensus protocol was introduced in \cite{MaxCSS, Molinari2, NoisyLinks}. Consider a network of $n$ agents with communication topology represented by an undirected graph $\mathcal{G} = (\mathcal{N}, \mathcal{E})$, where nodes represent agents and edges represent communication links.

\noindent Each agent $i \in \mathcal{N}$ is assigned an initial information state $x_{i}(0) \in \mathbb{R}$ at time step $t=0$. At each time step $t \in \mathbb{N}$, every agent updates its information state according to
\begin{align}
    x_i(t+1) = \max\bigl(x_i(t), \{x_j(t)\}_{j \in \mathcal{N}_i}\bigr).
    \label{eq:max_consensus}
\end{align}

\noindent Max-consensus is achieved when all agents converge to the maximum initial information state $x^{*}$ given by
\begin{align}
    \lim_{t \to \infty} x_{i}(t)  = x^{*} = \max_{i \in \mathcal{N}}(x_{i}(0))  \quad \forall i \in \mathcal{N}.
\end{align}

\noindent The time step $t_{\text{max}}$ required to converge to $x^{*}$ is denoted as the consensus convergence time. If consensus is achieved, it implies that
\begin{align}
\exists t_{\text{max}}: x_{i}(t) = x^{*} \quad  \forall t \geq t_{\text{max}}, \, \forall i \in \mathcal{N}.
\label{eq:k}
\end{align}

%% file: consensus.tex
The standard max-consensus protocol was extended to unreliable communication links in \cite{bernoulli,Katha}. We model link failures using a binary random variable $B_{ij}(t) \in \{0,1\}$ for edge $(i,j) \in \mathcal{E}$, where at time step $t$
\begin{align}
    B_{ij}(t) = \begin{cases}
        0 \quad \text{link is active (no failure)}\\
        1 \quad \text{link failure}.
    \end{cases}
\end{align}

\noindent Each link fails independently with probability $p_{ij} \in [0,1)$ such that $B_{ij}(t)$ follows a Bernoulli distribution with
\begin{align}
\begin{aligned}
    \Pr(B_{ij}(t) = 1) &= p_{ij}, \\
    \Pr(B_{ij}(t) = 0) &= 1 - p_{ij}.
\end{aligned}
\end{align}
\noindent Consequently, the transmission delay $X_{ij}$, defined as 
\mbox{$X_{ij} = \min\{t \geq 1 \mid B_{ij}(t) = 0\}$}, 
follows a geometric distribution, 
as introduced in Section~\ref{sec:GEO}.

At each time step, node $i$ can only receive information from neighbors whose links remain active, defining the effective neighboring set
\begin{align}
    \mathcal{N}_{i}^{\text{eff}}(t) = \{j \in \mathcal{N}_{i} \mid B_{ij}(t) = 0\}.
    \label{eq:N_i_eff}
\end{align}

Based on \eqref{eq:N_i_eff}, the max-consensus protocol under Bernoulli link failures is given by \cite{Katha}:
\begin{align}
    x_{i}(t+1) = \max \bigl(x_{i}(t), \{x_{j}(t)\}_{j \in \mathcal{N}_{i}^{\text{eff}}(t)}\bigr).
    \label{cons_new}
\end{align}

Let $t_{\text{max}}^{\text{eff}}$ denote the effective consensus convergence time, defined as the number of time steps required to reach max-consensus under probabilistic link failures. Since link failures delay information propagation, it follows that $t_{\text{max}}^{\text{eff}} \geq t_{\text{max}}$.

%% file: convergence.tex
\subsubsection{Ideal Communication Links}
We first review convergence guarantees for max-consensus under error-free communication. As established in \cite{MaxCSS, Molinari2}, the max-consensus convergence time over a strongly connected, time-invariant graph satisfies
\begin{align}
	t_{\max} \leq D,
	\label{eq:e_D_1}
\end{align}
\noindent where $D$ denotes the network diameter. This bound applies to arbitrary initial conditions.

When the source node $i^* \in \mathcal{N}$ holding the maximum initial 
value $x^*$ is known, the convergence time equals the eccentricity 
$e(i^*)$, because the maximum value propagates along shortest paths from 
$i^*$ to all other nodes. Thus, the bound in \eqref{eq:e_D_1} can be improved as
\begin{align}
	t_{\max} = e(i^*) \leq D.
\end{align}

\subsubsection{Unreliable Communication Links}
Under probabilistic link failures, the authors of~\cite{randomNetwork1} proved that 
max-consensus converges in four probabilistic senses (almost surely, in 
probability, in expectation, and mean square) if and only if the expected 
graph is strongly connected. Moreover, the authors of~\cite{bernoulli} derived 
an asymptotic upper bound on the expected convergence time as follows
\begin{align}
    \mathbb{E}[t_{\max}^{\mathrm{eff}}] \leq \frac{D}{1-p_{\max}}, 
    \quad p_{\max} = \max_{(i,j) \in \mathcal{E}} p_{ij}.
    \label{eq:golfar_bound}
\end{align}
Again, when the source node $i^*$ is known, applying $e(i^*) \leq D$ yields a tighter bound by replacing $D$ with $e(i^*)$ in~\eqref{eq:golfar_bound}. 

However, these results have two key limitations:
\begin{itemize}
    \item Using the worst-case link failure probability 
          $p_{\max}$ in \eqref{eq:golfar_bound} may overestimate the true expected convergence time.
    \item No existing method compute the \ac{PMF} and \ac{CDF} of $t_{\max}^{\mathrm{eff}}$, 
limiting the ability to assess deadline guarantees $\Pr(t_{\max}^{\mathrm{eff}} \leq \tau)$ 
for a given deadline $\tau$.
\end{itemize}

The remainder of this paper addresses these limitations by proposing the \ac{LiFE-CD} algorithm.

%% file: problem_new.tex
\subsection{Problem Formulation and Delay Modeling}
\label{ssec:problem}
The bound in~\eqref{eq:golfar_bound} treats the network as a uniform 
delay chain, relying solely on the diameter $D$ and worst-case link 
failure probability $p_{\max}$. To move beyond this asymptotic upper 
bound on the expected convergence time and compute the full convergence 
time distribution, three fundamental aspects must be addressed: (i) how 
delays accumulate differently depending on whether information propagates 
sequentially (unicast) or in parallel (broadcast), (ii) how these delay compositions lead to non-geometric distributions, and (iii) how the composed delay distributions can be characterized using shortest-path spanning trees. The following subsections address each aspect, forming the 
building blocks of \ac{LiFE-CD}.

\subsubsection{Unicast vs. Broadcast Transmission}
\begin{figure}[t]
\centering
\subfloat[]{\label{fig:uni}%
    \begin{tikzpicture}[
        node/.style={circle, draw, very thick, minimum size=10mm, font=\sffamily\large, fill=blue!8!teal!8!},
        source/.style={circle, draw, very thick, minimum size=10mm, font=\sffamily\large, fill=blue!25!teal!25},
        arrow/.style={<->, >={Stealth[length=2.5mm]}, thick},
        scale=0.75, transform shape
    ]
        \node[source] (1) at (0, 0) {1};
        \node[node] (2) at (2, 0) {2};
        \node[node] (dots) at (4, 0) {$\cdots$};
        \node[node] (n) at (6, 0) {$n$};
        \draw[arrow, <->, very thick] (1) -- (2);
        \draw[arrow, <->, very thick] (2) -- (dots);
        \draw[arrow, <->, very thick] (dots) -- (n);
    \end{tikzpicture}%
}
\hfill
\subfloat[]{\label{fig:broad}%
    \begin{tikzpicture}[
        node/.style={circle, draw, very thick, minimum size=10mm, font=\sffamily\large, fill=blue!8!teal!8!},
        source/.style={circle, draw, very thick, minimum size=10mm, font=\sffamily\large, fill=blue!25!teal!25},
        arrow/.style={<->, >={Stealth[length=2.5mm]}, thick},
        scale=0.75, transform shape
    ]
        \node[source] (1) at (3, 1.5) {1};
        \node[node] (2) at (0, 0) {2};
        \node[node] (3) at (2, 0) {3};
        \node[node] (dots) at (4, 0) {$\cdots$};
        \node[node] (n) at (6, 0) {$n$};
        \draw[arrow, <->, very thick] (1) -- (2);
        \draw[arrow, <->, very thick] (1) -- (3);
        \draw[arrow, <->, very thick] (1) -- (dots);
        \draw[arrow, <->, very thick] (1) -- (n);
    \end{tikzpicture}%
}
\caption{Transmission scenarios with node~$1$ as the source node. (a)~Unicast. (b)~Broadcast.}
\label{fig:unicast_broadcast}
\end{figure}
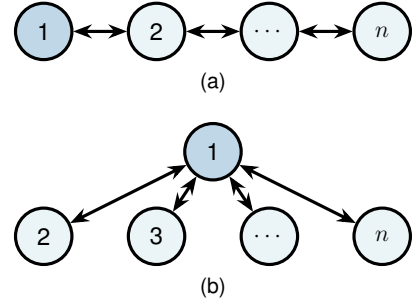

The bound in~\eqref{eq:golfar_bound} assumes propagation as a sequential transmission chain of length $D$, assigning each hop the worst-case expected delay based on $p_{\max}$. This ignores that information may propagate via branching 
nodes and discards the topology-dependent heterogeneity of the individual link failure probabilities $p_{ij}$. Instead,  information propagation in max-consensus can be decomposed into two fundamental transmission modes, as illustrated in Fig.~\ref{fig:unicast_broadcast}.

Unicast transmission (Fig.~\ref{fig:uni}): Information propagates sequentially along a path. Node $v_1$ transmits to node $v_2$, and upon successful reception, node $v_2$ forwards to node $v_3$ until node $v_n$ receives the information. Due to probabilistic link failures, each transmission attempt may fail, 
introducing a random delay $X_{v_k v_{k+1}}$ on the 
link $(v_k, v_{k+1})$ for $k = 1, \ldots, n-1$. The total delay is 
$Z_{\text{sum}}~=~\sum_{k=1}^{n-1}~X_{v_k v_{k+1}}$. By linearity of expectation we have
\begin{align}
    \mathbb{E}[Z_{\text{sum}}] = \sum_{k=1}^{n-1} \mathbb{E}[X_{v_k v_{k+1}}],
    \label{eq:unicast_linear}
\end{align}

\noindent meaning that the expected total delay equals the sum of the individual expected link delays. The bound~\eqref{eq:golfar_bound} can be seen as a special case of~\eqref{eq:unicast_linear}, where the chain has length
$D$ and all individual delays are replaced by the worst-case delay based on $p_{\max}$.

Broadcast transmission (Fig.~\ref{fig:broad}): A node simultaneously transmits to multiple neighbors. Consider node $v_1$ broadcasting to its children $v_2, \ldots, v_n$ 
with delays $X_{v_1 v_2}, \ldots, X_{v_1 v_n}$. The effective delay 
is $Z_{\parallel}~=~\max(X_{v_1 v_2}, \ldots, X_{v_1 v_n})$. A straightforward approach to estimate $\mathbb{E}[Z_{\parallel}]$ is considering the worst case, i.e., the maximum expected delay, similar to the approach in~\eqref{eq:golfar_bound}. However, this underestimates $\mathbb{E}[Z_{\parallel}]$ as can be seen in Jensens's inequality:
\begin{align}
    \mathbb{E}[Z_{\parallel}] \geq \max(\mathbb{E}[X_{v_1 v_2}], 
    \ldots, \mathbb{E}[X_{v_1 v_n}]).
    \label{eq:broadcast_jensen}
\end{align}

In contrast, the bound~\eqref{eq:golfar_bound} ignores broadcast transmissions and purely accounts for sequential unicast chains with worst-case 
delay $p_{\max}$, yielding an upper bound. Therefore, a correct analysis must explicitly distinguish between 
unicast and broadcast transmission modes. The following subsection 
characterizes how each transmission mode transforms the delay distributions, 
forming the basis for the exact computation of the convergence 
time distribution in \ac{LiFE-CD}.

\subsubsection{Non-Geometric Delay Distributions}
While individual links exhibit geometrically distributed delays, their compositions through sums (unicast) and maxima (broadcast) do not remain geometric. Consequently, computing the exact convergence time distribution requires explicitly tracking distribution evolution through convolution of the \ac{PMF}s and products of the \ac{CDF}s for both unicast and broadcast.

For unicast transmission over a path of $n-1$ links with delays $X_{v_1 v_2}, \ldots, X_{v_{n-1} v_n}$, the total delay $Z_{\text{sum}}~=~\sum_{k=1}^{n-1} X_{v_k v_{k+1}}$ has a \ac{PMF} given by the convolution
\begin{align}
    f_{Z_{\text{sum}}}(k) = f_{X_{v_1 v_2}} \ast f_{X_{v_2 v_3}} \ast \cdots \ast f_{X_{v_{n-1} v_n}}(k).
    \label{eq:sum}
\end{align}
The CDF $F_{Z_{\text{sum}}}(k)$ is obtained from \eqref{eq:sum} via \eqref{eq:CDF}, and the expected value via \eqref{eq:EV}.

For broadcast transmissions from a node to its $n-1$ children with delays $X_{v_1 v_2}, \ldots, X_{v_1 v_{n}}$, the maximum delay $Z_{\parallel}~=~\max(X_{v_1 v_2}, \ldots, X_{v_{1} v_n})$ has \ac{CDF} given by the pointwise product
\begin{align}
    F_{Z_{\parallel}}(k) = \prod_{i=1}^{n-1} F_{X_{v_i v_{i+1}}}(k).
    \label{eq:prod}
\end{align}
The \ac{PMF} $f_{Z_{\parallel}}(k)$ follows from \eqref{eq:prod} via \eqref{eq:PMF}, and the expected value via \eqref{eq:EV}.

By recursively applying the distribution-tracking operations \eqref{eq:sum} and \eqref{eq:prod} for a given network topology, our approach preserves the full probability distribution of the consensus convergence time, enabling expected value calculation and deadline probability assessment.

\subsubsection{Network Topology via Shortest-Path Trees}
\label{sec:shortest_path}
For acyclic networks, the graph has a tree structure with unique paths between any two nodes. For cyclic networks, however, multiple paths may exist between two nodes, and information propagates along whichever path completes first. Consider $n$ independent paths $P_1, \ldots, P_n$  with total delays $X_{P_1},...,X_{P_n}$.  However, explicitly tracking all path combinations is computationally prohibitive due to exponential path growth and edge-sharing dependencies, preventing direct recursive application of \eqref{eq:sum} and \eqref{eq:prod}.

To address this, we approximate the cyclic network by its shortest-path spanning tree, which removes cycles in the graph by considering minimum delay paths only. As a result, an upper bound on the consensus convergence time is obtained, which can be seen by Jensen's inequality as follows:
\begin{align}
    \mathbb{E}[\min(X_{P_1},...,X_{P_n})] \leq \min(\mathbb{E}[X_{P_1}],..., \mathbb{E}[X_{P_n}]).
    \label{eq:min_inequality}
\end{align}

Constructing the shortest-path spanning tree enables recursive application of~\eqref{eq:sum} and~\eqref{eq:prod} for cyclic graphs, yielding tight upper bounds on the consensus convergence time, c.f. Section \ref{sec:evaluation}.

As an example, Fig.~\ref{fig:cyclic_network} illustrates a cyclic network where information from source node~$1$ to node~$5$ can propagate via node~$3$ or node~$4$, creating path redundancy. For this network, two possible trees exist (solid and dashed arrows), with edges $e_{12}$, $e_{23}$, and $e_{24}$ common to both, while the selection of $(3,5)$ or $(4,5)$ depends on $p_{ij}$.

To obtain a tractable graph topology, we construct a shortest-path spanning tree rooted at the known source $i^*$. By selecting only the minimum expected-delay path to each node, this tree:
    
\begin{itemize}
    \item Eliminates cyclic redundancy while preserving the fastest propagation routes,
    \item Provides a tree structure enabling recursive application of \eqref{eq:sum} and \eqref{eq:prod}.
\end{itemize}

Together with the unicast/broadcast model and non-geometric delay composition, the shortest-path tree construction forms the third building block of the \ac{LiFE-CD} algorithm presented in Section~\ref{ssec:system}.

%% file: system_new.tex
\subsection{Algorithm Description}
\label{ssec:system}
Building on the delay modeling framework introduced in Section~\ref{ssec:problem}, where we discussed unicast/broadcast transmission modes, delay distribution calculation, and shortest-path tree construction for cyclic networks, we now present the \ac{LiFE-CD} algorithm, which is summarized in Algorithm~\ref{alg:LiFE-CT}. The objective of \ac{LiFE-CD} is to compute the full probability distribution of the consensus time for the max-consensus protocol under probabilistic link failures.

In contrast to existing analytical approaches, which derive bounds on the expected convergence time \cite{bernoulli}, \ac{LiFE-CD} computes the complete \ac{PMF} and \ac{CDF} of the consensus time. These provide valuable insights into the statistics of the consensus convergence time, e.g., expected value and deadline satisfaction probabilities. For acyclic networks, the resulting distribution is exact. For cyclic topologies, \ac{LiFE-CD} constructs a shortest-path spanning tree, yielding a tight upper bound on the consensus convergence time.

The algorithm takes as input the weighted adjacency matrix
\begin{align}
    W = [w_{ij}], \quad w_{ij} = p_{ij} \in (0,1)
\end{align}

\noindent of the underlying graph $\mathcal{G} = (\mathcal{N}, \mathcal{E}, W)$, where $p_{ij}$ denotes the link failure probability between nodes $i$ and $j$, and the source node $i^*$ holding the maximum initial information state $x_{i^*}(0) = x^*$. The output consists of the PMF $f_Z$ and CDF $F_Z$ of the consensus time random variable $Z$.

The \ac{LiFE-CD} algorithm is partitioned in four main phases:
\begin{enumerate}
    \item Graph preprocessing and cycle elimination (lines~1--3),
    \item Initialization of geometric distributions (lines~4--10),
    \item Iterative network reduction via unicast and broadcast operations (lines~11--31), and
    \item Final distribution composition (lines~33--39).
\end{enumerate}
\noindent The core principle (phase 3) is to recursively reduce the network topology via unicast and broadcast compositions until a single distribution remains.

\subsubsection*{Phase 1: Graph Preprocessing and Cycle Elimination}
To enable systematic distinction between unicast and broadcast transmission modes and to obtain a well-defined acyclic information propagation structure, we construct a shortest-path spanning tree as established in Section~\ref{sec:shortest_path}. Let $\mathcal{G} = (\mathcal{N}, \mathcal{E}, W)$ denote the original communication graph. For cyclic networks, we apply Dijkstra's algorithm~\cite{Dijkstra} to construct a shortest-path spanning tree $\mathcal{T} = (\mathcal{N}, \mathcal{E}_T, W)$ rooted at the source node $i^*$. This algorithm removes all cycles while selecting the shortest paths  $P_{i^*j}$ from the source to each destination node $j$, given by the set
\begin{equation}
    \mathcal{P}_{i^*,\text{SPT}} = \{ P_{i^*j} \mid j \in \mathcal{N},\ j \neq i^* \},
\end{equation}
where each path $P_{i^*j}$ is unique due to the tree structure. 

Among all root-to-node paths in $\mathcal{P}_{i^*,\text{SPT}}$, only those terminating at leaf nodes contribute independently to the consensus time. Paths ending at non-leaf nodes are subpaths of longer paths and do not require separate tracking. Removing all such redundant subpaths yields the set of critical paths:
\begin{align}
\mathcal{P}_{i^*,\text{critical}} = 
\left\{ P_{i^* j} \in \mathcal{P}_{i^*,\text{SPT}} \;\middle|\;
\nexists\, P_{i^* k} \in \mathcal{P}_{i^*,\text{SPT}},\ k \neq j,\right. \nonumber \\
\left. P_{i^* j} \sqsubseteq P_{i^* k} \right\}.
\label{eq:critical_def}
\end{align}
Critical paths represent independent information flows from the source to leaf nodes, jointly determining the consensus time and enabling unicast and broadcast reduction operations in phase 3.

\subsubsection*{Phase 2: Geometric Distribution Initialization}
The iterative reduction process starts at the leaf nodes of the 
critical paths. In the spanning tree $\mathcal{T}$, leaf nodes are 
those with degree $\deg(j) = 1$. Let
\begin{equation}
    \mathcal{N}_{i^*,\text{leaf}} =
    \left\{ j \in \mathcal{N} \setminus \{i^*\} \;\middle|\;
    \deg(j) = 1,\ P_{i^* j} \in 
    \mathcal{P}_{i^*,\text{critical}} \right\}
\end{equation}
denote the set of leaf nodes associated with the critical paths. For 
each $j \in \mathcal{N}_{i^*,\text{leaf}}$, let 
$q_j = \operatorname{parent}(j)$ denote its parent node in 
$\mathcal{T}$, with edge $(q_j,  j) \in \mathcal{E}_\mathcal{T}$. Since each node has a unique parent in 
$\mathcal{T}$, the link $(q_j, j)$ is uniquely determined by $j$. We denote by $X_{q_{j} j}$ the delay random variable from node $q_j$ to node $j$. At initialization, $X_{q_j j}$ follows a geometric distribution on link 
$(q_j, j)$ with failure probability $p_{q_{j} j}$, as each transmission 
attempt fails independently with probability $p_{q_j j}$. The corresponding 
\ac{PMF} and \ac{CDF} are given by~\eqref{eq:PMF_geometric} 
and~\eqref{eq:CDF_geometric}, respectively. The initial \ac{CDF}s 
constitute the set
\begin{equation}
    \mathcal{F}_{\text{leaf}} =
    \left\{ F_{X_{q_j j}} \;\middle|\; j \in \mathcal{N}_{i^*,\text{leaf}} \right\}.
\end{equation}
Starting from these geometric distributions, the algorithm recursively applies unicast and broadcast reductions to compute the final consensus time distribution.

\subsubsection*{Phase 3: Iterative Network Reduction}
The network is iteratively reduced until a single distribution remains. In each iteration, the longest remaining critical paths are processed first. Let
\begin{equation}
    \ell_{\max} = \max_{j \in \mathcal{N}_{i^*,\text{leaf}}}
    \left| \mathcal{E}(P_{i^* j}) \right|
\end{equation}
denote the maximum path length, and let
\begin{equation}
    \mathcal{L}_{\max} =
    \left\{ P_{i^* j} \in \mathcal{P}_{i^*,\text{critical}} \;\middle|\;
    \left| \mathcal{E}(P_{i^* j}) \right| = \ell_{\max} \right\}
\end{equation}
be the set of critical paths corresponding to this length.

Paths in $\mathcal{L}_{\max}$ are grouped according to the parent nodes $q_j$ of the leaf nodes $j \in \mathcal{N}_{i^*,\text{leaf}}$, yielding disjoint subsets 
\begin{equation}
\mathcal{L}_{\max,q_j} = \left\{ P_{i^{*}j} \in \mathcal{L}_{\max} \;\middle|\; \operatorname{parent(j)} = q_j \right\}.
\end{equation}

For each group, the transmission mode is determined by its cardinality as follows.

\paragraph{Unicast Reduction}
If $|\mathcal{L}_{\max,q_{j}}| = 1$, parent node $q_j$ has a single child, indicating unicast transmission. Let $q_j^{(2)} = \text{parent}(q_j)$ denote the parent of $q_j$, with edge $(q_j^{(2)}, q_j) \in \mathcal{E}_T$ and link failure probability $p_{q_j^{(2)} q_j}$. The transmission delay $X_{q_j^{(2)} q_j}$ over this link follows a geometric distribution with \ac{PMF} $f_{X_{q_j^{(2)} q_j}}$ and \ac{CDF} $F_{X_{q_j^{(2)} q_j}}$ as given in \eqref{eq:PMF_geometric} and \eqref{eq:CDF_geometric}, respectively.

For unicast transmission, delays accumulate additively via convolution. The combined \ac{CDF} is computed as
\begin{equation}
    F_{X_{q_j^{(2)} q_j}} \gets \operatorname{sum}(F_{X_{q_j j}}, F_{X_{q_j^{(2)} q_j}}),
\end{equation}
where the $\operatorname{sum}$ operator: (i) converts the input \ac{CDF}s to \ac{PMF}s via \eqref{eq:PMF}, (ii) convolves the \ac{PMF}s as $f_{\text{sum}} = f_{X_{q_j j}} \ast f_{X_{q_j^{(2)} q_j}}$ according to \eqref{eq:sum}, and (iii) converts the resulting \ac{PMF} back to a \ac{CDF} via \eqref{eq:CDF}.  

The resulting \ac{CDF} $F_{X_{q_j^{(2)} q_j}}$ replaces $F_{X_{q_j j}}$ in $\mathcal{F}_{\text{leaf}}$, the edge $(q_j, j)$ is removed from $\mathcal{E}_T$, and $\mathcal{P}_{i^{*},\text{critical}}$ and $\mathcal{N}_{i^{*},\text{leaf}}$ are updated accordingly.

\paragraph{Broadcast Reduction}
If $|\mathcal{L}_{\max,q_j}| > 1$, the parent node $q_j$ transmits via broadcast to its children. We identify the child nodes $j_1,...,j_m$, and the \ac{CDF}s $F_{X_{q_j j_1}},...,F_{X_{q_j j_m}} \in \mathcal{F}_\text{leaf}$. 

Since, for broadcast transmission, the resulting delay is determined by the maximum delay, the combined \ac{CDF} is computed as
\begin{equation}
    F_{X_{q_j j_1}} \gets \operatorname{max}(F_{X_{q_j j_1}}, \ldots, F_{X_{q_j j_m}}),
\end{equation}
computed by pointwise multiplication  of the \ac{CDF}s as in \eqref{eq:prod}. The resulting $F_{X_{q_j j_1}}$ is updated in $\mathcal{F}_{\text{leaf}}$, while $F_{X_{q_j j_2}}, \ldots, F_{X_{q_j j_m}}$ are removed, edges $(q_j, j_2), \ldots, (q_j, j_m)$ are removed from $\mathcal{E}_T$, and $\mathcal{P}_{i^{*},\text{critical}}$ and $\mathcal{N}_{i^{*},\text{leaf}}$ are updated accordingly.

After each unicast or broadcast reduction step, the updated sets 
$\mathcal{F}_{\text{leaf}}$, $\mathcal{P}_{i^{*},\text{critical}}$, 
and $\mathcal{N}_{i^{*},\text{leaf}}$ serve as input to the 
subsequent iteration of the network reduction.

\subsubsection*{Phase 4: Termination and Final Distribution}
The reduction process terminates when either a single unicast path remains ( \( |\mathcal{N}_{i^{*},\text{leaf}}| = 1 \) and \( |\mathcal{E}(P_{i^{*}j})| = 2 \)) or all remaining leaf nodes are directly connected to the source node ($|\mathcal{N}_{i^*,\text{leaf}}| > 1$ and $|\mathcal{E}(P_{i^{*}j})| = 1 $) indicating a broadcast transmission mode. In both cases, a final sum or max operation is performed, yielding the final \ac{PMF} $f_Z$ and \ac{CDF} $F_Z$ of the consensus time $Z$, which can be used for further statistical evaluations, e.g., based on the expected value $\mathbb{E}[Z]$ or the deadline satisfaction probability $\Pr(Z \le \tau)$ for a given deadline $\tau$.

\begin{algorithm}[!ht]
\caption{\ac{LiFE-CD}: Link Failure Effects on Consensus Distributions}
\label{alg:LiFE-CT}
\begin{algorithmic}[1]
\Require $W = [p_{ij}]$, source node $i^*$
\Ensure \ac{PMF} $f_Z$, \ac{CDF} $F_Z$
\State \textbf{Phase 1: Graph Preprocessing}
\State Compute shortest-path tree $\mathcal{P}_{i^*,\text{SPT}}$ rooted at $i^*$ 
\State Construct critical paths $\mathcal{P}_{i^*,\text{critical}}$ by removing subpaths
\State \textbf{Phase 2: Distribution Initialization}
\State Identify leaf nodes $\mathcal{N}_{i^*,\text{leaf}}$ from $\mathcal{P}_{i^*,\text{critical}}$
\For{$j \in \mathcal{N}_{i^*,\text{leaf}}$}
    \State $q_j \gets \text{parent}(j)$
    \State $F_{X_{q_j j}} \gets$ geometric \ac{CDF} with $p_{q_j j}$
\EndFor
\State $\mathcal{F}_{\text{leaf}} \gets \{F_{X_{q_j j}} \mid j \in \mathcal{N}_{i^*,\text{leaf}}\}$
\State \textbf{Phase 3: Iterative Reduction}
\While{not terminal condition} \Comment{See Phase 4}
    \State $\ell_{\max} \gets \max_{j \in \mathcal{N}_{i^*,\text{leaf}}} |\mathcal{E}(P_{i^*j})|$
    \State $\mathcal{L}_{\max} \gets \{P_{i^*j} \in \mathcal{P}_{i^*,\text{critical}} \mid |\mathcal{E}(P_{i^*j})| = \ell_{\max}\}$
     \State $\mathcal{L}_{\max,q_j} \gets \left\{ P_{i^{*}j} \in \mathcal{L}_{\max} \;\middle|\; \text{parent}(j) = q_j \right\}$
    \For{each group $\mathcal{L}_{\max,q_j}$}
        \If{$|\mathcal{L}_{\max,q_j}| = 1$} \Comment{Unicast}
            \State Identify $q_j^{(2)} = \text{parent}(q_j)$
            \State $F_{X_{q_j^{(2)} q_j}} \gets$ geometric \ac{CDF} with $p_{q_j^{(2)}
             q_j}$
            \State $F_{X_{q_j^{(2)} q_j}} \gets \operatorname{sum}(F_{X_{q_j j}}, F_{X_{q_j^{(2)} q_j}})$ 
            \State Replace $F_{X_{q_j j}}$ in $\mathcal{F}_{\text{leaf}}$ with $F_{X_{q_j^{(2)} q_j}}$
            \State Remove $(q_j, j)$ from $\mathcal{E}_T$
        \Else \Comment{Broadcast}
            \State Identify leaf nodes $j_1, \ldots, j_m$ in $\mathcal{L}_{\max,q_j}$
            \State $F_{X_{q_j j_1}} \gets \operatorname{max}(F_{X_{q_j j_1}}, \ldots, F_{X_{q_j j_m}})$
            \State Update $F_{X_{q_j j_1}}$ in $\mathcal{F}_{\text{leaf}}$
            \State Remove $F_{X_{q_j j_2}}, \ldots, F_{X_{q_j j_m}}$ from $\mathcal{F}_{\text{leaf}}$
            \State Remove  $(q_j, j_2), \ldots, (q_j, j_m)$ from $\mathcal{E}_T$
        \EndIf
    \EndFor
    \State Update $\mathcal{P}_{i^*,\text{critical}}$ and $\mathcal{N}_{i^*,\text{leaf}}$
\EndWhile
\State \textbf{Phase 4: Final Composition}
\If{$|\mathcal{N}_{i^*,\text{leaf}}| = 1$ and $|\mathcal{E}(P_{i^{*}j})| = 2 $}
    \State repeat lines 18--20: $F_Z \gets \operatorname{sum}(F_{X_{q_j j}}, F_{X_{q_j^{(2)} q_j}})$ 
\ElsIf {$|\mathcal{N}_{i^{*},\text{leaf}| > 1}$ and $|\mathcal{E}(P_{i^{*}j})| = 1$}
    \State repeat lines 24--25: $F_Z \gets \operatorname{max}(F_{X_{q_j j_1}}, \ldots, F_{X_{q_j j_m}})$ 
\EndIf 
\State Compute \ac{PMF} $f_Z$ 
\State \Return $f_Z$, $F_Z$
\end{algorithmic}
\end{algorithm}

%% file: example.tex
\subsubsection{Graph Preprocessing and Distribution Initialization}
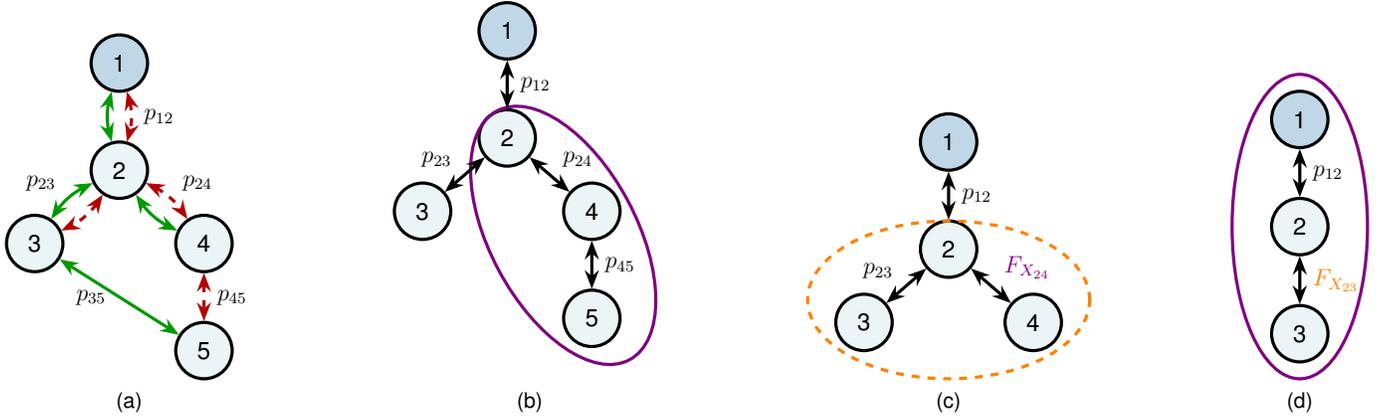
\begin{figure*}[t]
\centering
\subfloat[]{\label{fig:cyclic_network}%
    \begin{tikzpicture}[
        node/.style={circle, draw, very thick, minimum size=10mm, font=\sffamily\large, fill=blue!8!teal!8!},
        source/.style={circle, draw, very thick, minimum size=10mm, font=\sffamily\large, fill=blue!25!teal!25},
        arrow/.style={<->, >={Stealth[length=2.5mm]}, thick},
        label/.style={font=\large, text=black},
        scale=0.75, transform shape
    ]
        \node[source] (1) at (0, 2.4) {1};
        \node[node] (2) at (0, 0.5) {2};
        \node[node] (3) at (-1.5, -0.8) {3};
        \node[node] (4) at (1.5, -0.8) {4};
        \node[node] (5) at (1.5, -2.7) {5};
        \draw[arrow, <->, green!60!black, very thick] (1) to[bend right=15] (2);
        \draw[arrow, <->, green!60!black, very thick] (2) to[bend right=15] node[label, above left, xshift=-1mm] {$p_{23}$} (3);
        \draw[arrow, <->, green!60!black, very thick] (3) -- node[label, left, xshift=-1mm] {$p_{35}$} (5);
        \draw[arrow, <->, red!70!black, very thick, dashed] (1) to[bend left=15] node[label, right, xshift=1mm] {$p_{12}$} (2);
        \draw[arrow, <->, red!70!black, very thick, dashed] (2) to[bend left=15] node[label, above right, xshift=1mm] {$p_{24}$} (4);
        \draw[arrow, <->, red!70!black, very thick, dashed] (4) -- node[label, right, xshift=1mm] {$p_{45}$} (5);
        \draw[arrow, <->, green!60!black, very thick] (2) to[bend right=15] (4);
        \draw[arrow, <->, red!70!black, very thick, dashed] (2) to[bend left=15] (3);
    \end{tikzpicture}%
}
\hfill
\subfloat[]{\label{fig:J1}%
    \begin{tikzpicture}[
        node/.style={circle, draw, very thick, minimum size=10mm, font=\sffamily\large, fill=blue!8!teal!8!},
        source/.style={circle, draw, very thick, minimum size=10mm, font=\sffamily\large, fill=blue!25!teal!25},
        arrow/.style={<->, >={Stealth[length=2.5mm]}, thick},
        label/.style={font=\large, text=black},
        scale=0.75, transform shape
    ]
        \node[source] (1) at (0, 2.4) {1};
        \node[node] (2) at (0, 0.5) {2};
        \node[node] (3) at (-1.5, -0.8) {3};
        \node[node] (4) at (1.5, -0.8) {4};
        \node[node] (5) at (1.5, -2.7) {5};
        \draw[violet, very thick, rotate=28] (0.3, -1.55) ellipse (1.3cm and 2.5cm);
        \draw[arrow, <->, very thick] (1) -- node[label, right, xshift=1mm] {$p_{12}$} (2);
        \draw[arrow, <->, very thick] (2) -- node[label, above left, xshift=-1mm] {$p_{23}$} (3);
        \draw[arrow, <->, very thick] (2) -- node[label, above right, xshift=1mm] {$p_{24}$} (4);
        \draw[arrow, <->, very thick] (4) -- node[label, right, xshift=1mm] {$p_{45}$} (5);
    \end{tikzpicture}%
}
\hfill
\subfloat[]{\label{fig:J2}%
    \begin{tikzpicture}[
        node/.style={circle, draw, very thick, minimum size=10mm, font=\sffamily\large, fill=blue!8!teal!8!},
        source/.style={circle, draw, very thick, minimum size=10mm, font=\sffamily\large, fill=blue!25!teal!25},
        arrow/.style={<->, >={Stealth[length=2.5mm]}, thick},
        label/.style={font=\large, text=black},
        scale=0.75, transform shape
    ]
        \node[source] (1) at (0, 2.4) {1};
        \node[node] (2) at (0, 0.5) {2};
        \node[node] (3) at (-1.5, -0.8) {3};
        \node[node] (4) at (1.5, -0.8) {4};
        \draw[orange, very thick, dashed] (0, -0.4) ellipse (2.5cm and 1.4cm);
        \draw[arrow, <->, very thick] (1) -- node[label, right, xshift=1mm] {$p_{12}$} (2);
        \draw[arrow, <->, very thick] (2) -- node[label, above left, xshift=-1mm] {$p_{23}$} (3);
        \draw[arrow, <->, very thick] (2) -- node[label, above right, xshift=1mm, violet] {$F_{X_{24}}$} (4);
    \end{tikzpicture}%
}
\hfill
\subfloat[]{\label{fig:J3}%
    \begin{tikzpicture}[
        node/.style={circle, draw, very thick, minimum size=10mm, font=\sffamily\large, fill=blue!8!teal!8!},
        source/.style={circle, draw, very thick, minimum size=10mm, font=\sffamily\large, fill=blue!25!teal!25},
        arrow/.style={<->, >={Stealth[length=2.5mm]}, thick},
        label/.style={font=\large, text=black},
        scale=0.75, transform shape
    ]
        \node[source] (1) at (0, 2.4) {1};
        \node[node] (2) at (0, 0.5) {2};
        \node[node] (3) at (0, -1.4) {3};
        \draw[arrow, <->, very thick] (1) -- node[label, right, xshift=1mm] {$p_{12}$} (2);
        \draw[arrow, <->, very thick] (2) -- node[label, right, xshift=1mm, orange] {$F_{X_{23}}$} (3);
        \draw[violet, very thick] (0, 0.5) ellipse (1.2cm and 2.7cm);
    \end{tikzpicture}%
}
\caption{Network topologies for LiFE-CD algorithm analysis.
(a) Cyclic network with two possible shortest-path trees rooted at node~$1$ (green solid and red dashed lines).
(b)--(d) Iterative reduction of an acyclic network. Solid purple and dashed orange ellipses indicate sum and max operations, respectively.
(b) Original topology; (c) after sum operation for unicast; (d) after max operation for broadcast.}
\label{fig:network_topologies}
\end{figure*}

We demonstrate the \ac{LiFE-CD} algorithm adopting the network shown in Fig.~\ref{fig:J1}, with source node $i^* = 1$. Since the network is acyclic, we already have a shortest-path spanning tree. The critical paths and initial leaf nodes are given by
\begin{align}
    \mathcal{P}_{1,\text{critical}} &= \{P_{13}, P_{15}\} \label{eq:critical_1} \\
    \mathcal{N}_{1,\text{leaf}} &= \{3,5\}, \label{eq:leaf_1}
\end{align}
where  $P_{13} = (1,2,3)$ and $P_{15} = (1,2,4,5)$.

We initialize geometric \ac{CDF}s $F_{X_{23}}$ and $F_{X_{45}}$ for the leaf nodes with failure probabilities $p_{23}$ and $p_{45}$ and obtain the set of geometric \ac{CDF}s $\mathcal{F}_{\text{leaf}} = \{ F_{X_{23}}, F_{X_{45}}\}$.

\subsubsection{Iterative Reduction (Unicast)}
The longest path is $P_{15}$ with $\ell_{\max} = 3$. Since $\mathcal{L}_{\max,4} = \{P_{15}\}$ contains a single path, this is a unicast transmission. For the parent link $(2,4)$ with link failure probability $p_{24}$, the geometric \ac{CDF} $F_{X_{24}}$ is used to compute the sum operation:
\begin{align}
    F_{X_{24}} \gets \operatorname{sum}(F_{X_{45}}, F_{X_{24}}).
\end{align}

The set of leaf node distributions, critical paths, and the set of leaf nodes are updated to $\mathcal{F}_{\text{leaf}} = \{ F_{X_{23}}, F_{X_{24}}\}$, $\mathcal{P}_{1,\text{critical}} = \{P_{13}, P_{14}\}$ with $P_{14} = (1,2,4)$, and $\mathcal{N}_{1,\text{leaf}} = \{3,4\}$, respectively (Fig.~\ref{fig:J2}).

\subsubsection{Iterative Reduction (Broadcast)}
Now, $\ell_{\max} = 2$ with $\mathcal{L}_{\max,2} = \{P_{13}, P_{14}\}$. Since $|\mathcal{L}_{\max,1}| = 2$, this is a broadcast transmission, which applies the max operation
\begin{align}
    F_{X_{23}} \gets \operatorname{max}(F_{X_{23}}, F_{X_{24}})
\end{align}
and updates $\mathcal{F}_{\text{leaf}} = \{ F_{X_{23}}\}$, $P_{1,\text{critical}} = \{P_{13}\}$ with $P_{13} = (1,2,3)$, and $\mathcal{N}_{1,\text{leaf}} = \{3\}$ (Fig.~\ref{fig:J3}).

\subsubsection{Final Composition (Unicast)}
The terminal condition $|\mathcal{N}_{1,\text{leaf}}| = 1$ and $|\mathcal{E}(P_{13})| = 2$ is satisfied. Thus, the final sum operation is applied using the geometric \ac{CDF} $F_{X_{12}}$ with link failure probability $p_{12}$ for the link $(1,2)$ as follows
\begin{align}
    F_Z \gets \operatorname{sum}(F_{X_{23}}, F_{X_{12}}).
\end{align}
The \ac{PMF} $f_Z$ is obtained via \eqref{eq:PMF}.

%% file: eval2.tex
\subsection{Probability Distribution of Convergence Time}

The \ac{LiFE-CD} algorithm computes 
the full probability distribution of the consensus convergence time under link failures, providing rich 
information for system analysis and protocol design. In this section, the distributions computed by \ac{LiFE-CD} are compared with the Monte Carlo simulation results.

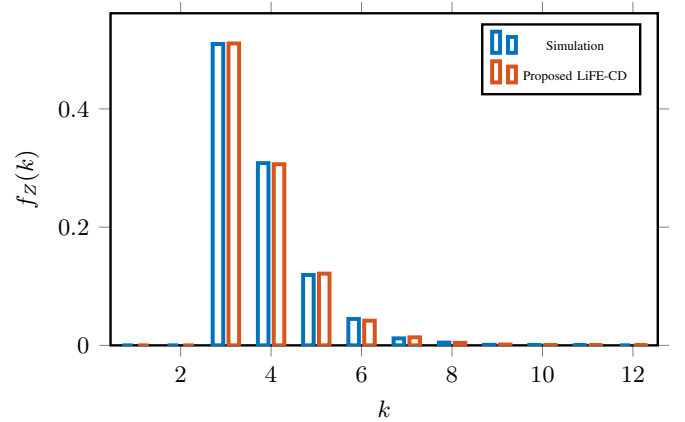
\begin{figure}[t]
\centering
\begin{tikzpicture}
\begin{axis}[
 width=\columnwidth,
    height=6cm,
    ybar,
    bar width=4pt,
    xlabel={$k$},
    ylabel={$f_{Z}(k)$},
    xmin=1, xmax=12,       
    enlarge x limits=0.05, 
    legend style={at={(0.97,0.97)}, anchor=north east},
    ymin=0,
    font=\small,
    grid=none,
    tick align=outside,
    line width=0.9pt,
    every axis plot/.append style={line width=1.5pt},
    legend style={
        font=\tiny,
      }
]

\addplot[
   color={rgb,1:red,0.00;green,0.45;blue,0.74},
] table [x=k, y=pmf_sim, col sep=comma] {data6.csv};
\addlegendentry{Simulation}

\addplot[
    color={rgb,1:red,0.85;green,0.33;blue,0.10},
] table [x=k, y=pmf_calc, col sep=comma] {data6.csv};
\addlegendentry{Proposed LiFE-CD}

\end{axis}
\end{tikzpicture}
\caption{PMF of the consensus convergence time $Z$: Simulation vs.\ LiFE-CD for the cyclic network in Fig. \ref{fig:cyclic_network}.}
\label{fig:pmf_acyclic}
\end{figure}

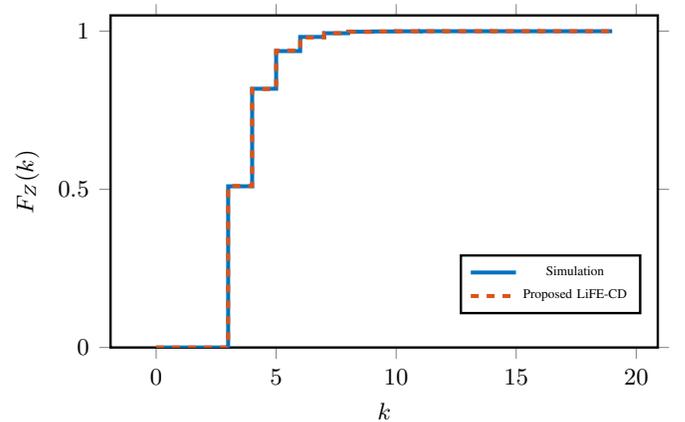
\begin{figure}[t]
\centering
\begin{tikzpicture}
\begin{axis}[
 width=\columnwidth,
    height=6cm,
    xlabel={$k$},
    ylabel={$F_Z(k)$},
    legend style={at={(0.97,0.1)}, anchor=south east},
    ymin=0, ymax=1.05,
    font=\small,
    grid=none,
    tick align=outside,
    line width=0.9pt,
    every axis plot/.append style={line width=1.5pt},
     legend style={
        font=\tiny,
      }
]

\addplot[
    const plot,
    color={rgb,1:red,0.00;green,0.45;blue,0.74},
    mark=none,
] table [x=k, y=cdf_sim, col sep=comma] {data7.csv};
\addlegendentry{Simulation}

\addplot[
    const plot,
    dashed,
    color={rgb,1:red,0.85;green,0.33;blue,0.10},
    mark=none,
] table [x=k, y=cdf_calc, col sep=comma] {data7.csv};
\addlegendentry{Proposed LiFE-CD}

\end{axis}
\end{tikzpicture}%
\caption{CDF of the consensus convergence time $Z$: Simulation vs.\ LiFE-CD for the acyclic network in Fig. \ref{fig:J1}.}
\label{fig:cdf_acyclic}
\end{figure}

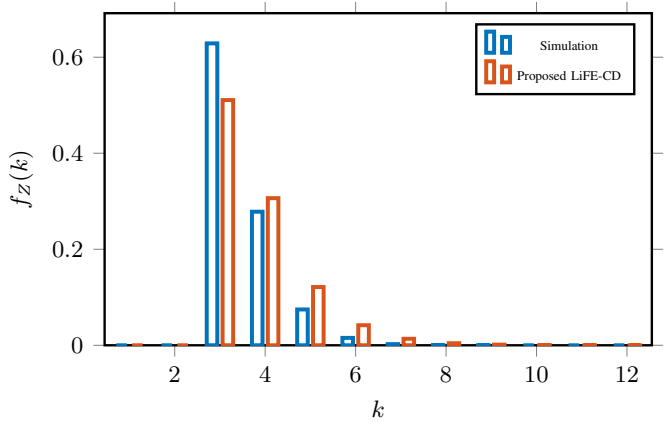
\begin{figure}[t]
\centering
\begin{tikzpicture}
\begin{axis}[
 	width=\columnwidth,
    height=6cm,
    ybar,
    bar width=4pt,
    xlabel={$k$},
    ylabel={$f_{Z}(k)$},
    xmin=1, xmax=12,       
    enlarge x limits=0.05, 
    legend style={at={(0.97,0.97)}, anchor=north east},
    ymin=0,
    font=\small,
    grid=none,
    tick align=outside,
    line width=0.9pt,
    every axis plot/.append style={line width=1.5pt},
    legend style={
        font=\tiny,
      }
]

\addplot[
   color={rgb,1:red,0.00;green,0.45;blue,0.74},
] table [x=k, y=pmf_sim, col sep=comma] {data4.csv};
\addlegendentry{Simulation}

\addplot[
    color={rgb,1:red,0.85;green,0.33;blue,0.10},
] table [x=k, y=pmf_calc, col sep=comma] {data4.csv};
\addlegendentry{Proposed LiFE-CD}

\end{axis}
\end{tikzpicture}
\caption{PMF of the consensus convergence time $Z$: Simulation vs.\ LiFE-CD for the cyclic network in Fig. \ref{fig:cyclic_network}.}
\label{fig:pmf_cyclic}
\end{figure}

\begin{figure}[t]
\centering
\begin{tikzpicture}
\begin{axis}[
 width=\columnwidth,
    height=6cm,
    xlabel={$k$},
    ylabel={$F_Z(k)$},
    legend style={at={(0.97,0.1)}, anchor=south east},
    ymin=0, ymax=1.05,
    font=\small,
    grid=none,
    tick align=outside,
    line width=0.9pt,
    every axis plot/.append style={line width=1.5pt},
     legend style={
        font=\tiny,
      }
]

\addplot[
    const plot,
    color={rgb,1:red,0.00;green,0.45;blue,0.74},
    mark=none,
] table [x=k, y=cdf_sim, col sep=comma] {data5.csv};
\addlegendentry{Simulation}

\addplot[
    const plot,
    dashed,
    color={rgb,1:red,0.85;green,0.33;blue,0.10},
    mark=none,
] table [x=k, y=cdf_calc, col sep=comma] {data5.csv};
\addlegendentry{Proposed LiFE-CD}

\end{axis}
\end{tikzpicture}%
\caption{CDF of the consensus convergence time $Z$: Simulation vs. LiFE-CD for the cyclic network in Fig. \ref{fig:cyclic_network}.}
\label{fig:cdf_cyclic}
\end{figure}
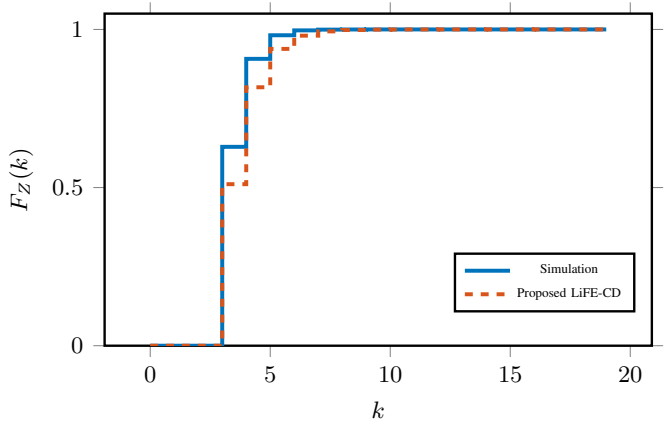

Figs.~\ref{fig:pmf_acyclic} and~\ref{fig:cdf_acyclic} show 
the \ac{PMF} and \ac{CDF} for the acyclic network in Fig. \ref{fig:J1}, 
while Figs.~\ref{fig:pmf_cyclic} and~\ref{fig:cdf_cyclic} 
show the corresponding results for the cyclic network in Fig. \ref{fig:cyclic_network}. The link failure probabilities were chosen for a representative example to $p_{12} = \qty{5}{\percent}$, $p_{23} = \qty{20}{\percent}$, $p_{24} = \qty{20}{\percent}$, $p_{45} = \qty{30}{\percent}$, and $p_{35} = \qty{60}{\percent}$.

\subsubsection{Acyclic Network}
For the acyclic network shown in Figs.~\ref{fig:pmf_acyclic} and~\ref{fig:cdf_acyclic}, the \ac{PMF} and \ac{CDF} 
obtained from Monte Carlo simulation coincide with the distributions across all values 
of $k$ computed by \ac{LiFE-CD}, confirming that \ac{LiFE-CD} outputs the exact 
probability distribution of the convergence time for 
acyclic topologies.

\subsubsection{Cyclic Network}
For the cyclic network, as shown in Figs.~\ref{fig:pmf_cyclic} and~\ref{fig:cdf_cyclic}, both the simulation and 
\ac{LiFE-CD} \ac{PMF} peak at $k = 3$; however, 
\ac{LiFE-CD} assigns lower probability mass at $k = 3$ 
and redistributes it toward $k \geq 4$. In other words, compared to \ac{LiFE-CD}, the simulated \ac{CDF} shows higher probabilities for smaller values of $k$, indicating that 
convergence occurs earlier in simulation than \ac{LiFE-CD} 
predicts. This is consistent with the upper-bound character 
of \ac{LiFE-CD} for cyclic topologies: the spanning-tree 
construction evaluates only the dominant shortest propagation path 
and does not account for path diversity, although alternative 
paths may enable faster convergence in individual simulation 
runs. Consequently, \ac{LiFE-CD} shifts probability mass 
toward larger values of $k$ relative to the empirical 
distribution, yet correctly bounds the expected value from 
above.

\subsubsection{Practical Relevance for System Design}
While the \ac{PMF} characterizes the shape and spread 
of the convergence time distribution, the \ac{CDF} is 
of direct practical relevance for deadline-aware protocol 
design. Given a reliability requirement $\tau \in (0,1)$, the minimum number 
of iterations $k^*$ guaranteeing that the consensus time 
$Z := t_{\max}^{\text{eff}}$ does not exceed $k^*$ with probability 
at least $\tau$ is obtained by
\begin{equation}
    k^*(\tau) = \min\bigl\{k \in \mathbb{N} \, | \, 
    \Pr(Z \leq k) \geq \tau \bigr\}.
    \label{eq:deadline}
\end{equation}
This enables a system designer to select a protocol timeout $k^*$ 
that guarantees max-consensus within $k^*$ iterations with 
confidence $\tau$, bounding the probability of non-convergence 
by $1 - \tau$. Probabilistic guarantees of this form are essential in 
real-time control and robotics applications, where controllers must 
enforce task completion with a specified confidence level while 
respecting temporal deadlines~\cite{Control2}.

%% file: eval1.tex
\subsection{Expected Convergence Time}

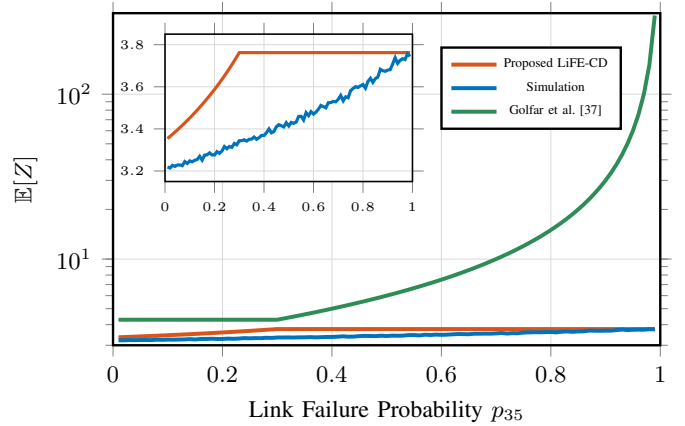
\begin{figure}[ht]
\centering
\begin{tikzpicture}
\begin{semilogyaxis}[
    name=mainplot,
    width=\columnwidth,
    height=6cm,
    xlabel={Link Failure Probability $p_{35}$},
    ylabel={$\mathbb{E}[Z]$},
    ymin=3, ymax=310,
    xmin=0, xmax=1,
    grid=major,
    grid style={line width=0.3pt, draw=gray!30},
    font=\small,
    tick align=outside,
    line width=0.9pt,
    every axis plot/.append style={line width=1.5pt},
    legend style={
        font=\tiny,
        at={(0.93,0.9)},
        anchor=north east,
    },
    legend entries={Proposed LiFE-CD, Simulation, Golfar et al. \cite{bernoulli}},
]

\addplot[color={rgb,1:red,0.85;green,0.33;blue,0.10}]
    table[x=u, y=calc, col sep=comma]{data.csv};

\addplot[color={rgb,1:red,0.00;green,0.45;blue,0.74}]
    table[x=u, y=sim, col sep=comma]{data.csv};

\addplot[color={rgb,1:red,0.18;green,0.55;blue,0.34}]
    table[x=u, y=golfar, col sep=comma]{data.csv};

\end{semilogyaxis}

\begin{axis}[
    name=inset,
    width=0.55\columnwidth,
    height=0.4\columnwidth,
   at={(rel axis cs:0.095,0.7)},
   anchor=north west,
    ymin=3.15, ymax=3.85,
    xmin=0, xmax=1,
    restrict y to domain=3.0:4.0,
    grid=major,
    grid style={line width=0.3pt, draw=gray!30},
    font=\tiny,
    tick align=outside,
    line width=0.6pt,
    every axis plot/.append style={line width=1.2pt},
    axis background/.style={fill=white},
    axis lines=box,
]

\addplot[color={rgb,1:red,0.85;green,0.33;blue,0.10}]
    table[x=u, y=calc, col sep=comma]{data.csv};

\addplot[color={rgb,1:red,0.00;green,0.45;blue,0.74}]
    table[x=u, y=sim, col sep=comma]{data.csv};

\addplot[color={rgb,1:red,0.18;green,0.55;blue,0.34}]
    table[x=u, y=golfar, col sep=comma]{data.csv};

\end{axis}

\end{tikzpicture}
\caption{Expected convergence time $\mathbb{E}[Z]$ over link failure probability $p_{35}$ for the cyclic network in Fig \ref{fig:cyclic_network}.}
\label{fig:convergence}
\end{figure}

\begin{figure}[ht]
\centering
\begin{tikzpicture}
\begin{semilogyaxis}[
    name=mainplot,
    width=\columnwidth,
    height=6cm,
    xlabel={Link Failure Probability $p_{45}$},
    ylabel={$\mathbb{E}[Z]$},
    ymin=3, ymax=300,
    xmin=0.01, xmax=1,
    grid=major,
    grid style={line width=0.3pt, draw=gray!30},
    font=\small,
    tick align=outside,
    line width=0.9pt,
    every axis plot/.append style={line width=1.5pt},
    legend style={
        font=\tiny,
        at={(0.35,0.95)},
        anchor=north east,
    },
    legend entries={Proposed LiFE-CD, Simulation, Golfar et al. \cite{bernoulli}},
]

\addplot[color={rgb,1:red,0.85;green,0.33;blue,0.10}]
    table[x=u, y=calc, col sep=comma]{data2.csv};

\addplot[dashed, color={rgb,1:red,0.00;green,0.45;blue,0.74}]
    table[x=u, y=sim, col sep=comma]{data2.csv};

\addplot[color={rgb,1:red,0.18;green,0.55;blue,0.34}]
    table[x=u, y=golfar, col sep=comma]{data2.csv};

\end{semilogyaxis}

\end{tikzpicture}
\caption{Expected convergence time $\mathbb{E}[Z]$ over link failure probability $p_{45}$ for the acyclic network in Fig \ref{fig:J1}.}
\label{fig:convergence2}
\end{figure}
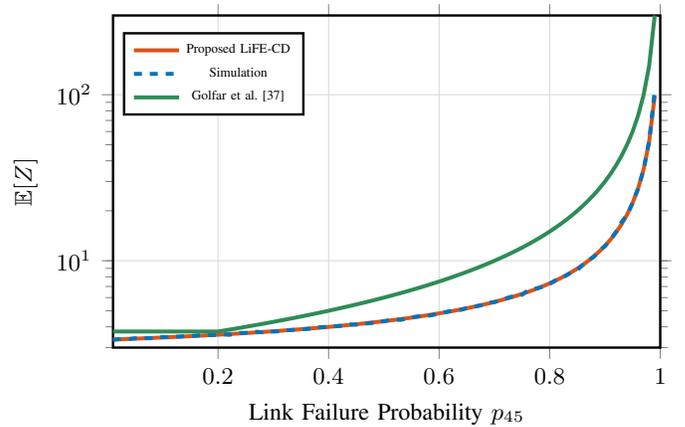

In this section, the performance of \ac{LiFE-CD} is evaluated against Monte Carlo simulation and the bound presented in~\cite{bernoulli} for both acyclic and cyclic network topologies with source node $i^* = 1$. For both network examples in Figs.~\ref{fig:cyclic_network} and~\ref{fig:J1}, the eccentricity of node $i^*$ equals the graph diameter, such that substituting $e(i^*)$ for $D$ in~\eqref{eq:golfar_bound} yields no additional tightening, and the bound presented in~\cite{bernoulli} is compared in its strongest form.

\subsubsection{Cyclic Network}
For the cyclic case, we consider the network topology of 
Fig.~\ref{fig:cyclic_network} with fixed link failure probabilities 
$p_{12} = \qty{5}{\percent}$, $p_{23} = \qty{20}{\percent}$, $p_{24} = \qty{20}{\percent}$, $p_{45} = \qty{30}{\percent}$
while $p_{35}$ is varied over the feasible interval $[0,1)$. 
Fig.~\ref{fig:convergence} shows the expected convergence time 
$\mathbb{E}[Z]$ as a function of $p_{35}$.

\ac{LiFE-CD} and the Monte Carlo simulation are in close 
agreement across the entire range of $p_{35}$, whereas 
the bound in~\cite{bernoulli} 
overestimates $\mathbb{E}[Z]$ by a factor of almost $8$ at 
$p_{35} = 90\%$ ($30$ vs.\ $3.76$), with the gap 
growing without bound as $p_{35} \to 1$. These results demonstrate that the proposed \ac{LiFE-CD} significantly improves the upper bound in~\cite{bernoulli}.

The constant $\mathbb{E}[Z]$ in Fig. \ref{fig:convergence} for \ac{LiFE-CD} is explained as follows. For $p_{35} < \qty{30}{\percent}$, \ac{LiFE-CD} computes $\mathbb{E}[Z]$ along 
the shortest-path tree indicated by the green path in 
Fig.~\ref{fig:cyclic_network}. At $p_{35} = p_{45} = \qty{30}{\percent}$, the 
expected cost of both green and red paths becomes equal, and for $p_{35} \geq \qty{30}{\percent}$ 
the dominant propagation path transitions to the red path in 
Fig.~\ref{fig:cyclic_network}, which is independent of $p_{35}$ 
and yields the constant value $\mathbb{E}[Z] = 3.76$.

As $p_{35}$ increases, the simulation results approach the 
\ac{LiFE-CD} bound of $\mathbb{E}[Z] = 3.76$. For $p_{35} = 1$, link $(3,5)$ 
is blocked, so the network effectively 
reduces to a single dominant propagation path, for which 
\ac{LiFE-CD} is exact. For $p_{35} < 1$, multiple 
paths remain simultaneously available, enabling faster convergence 
in individual runs and thus introducing a gap between the 
results obtained by simulation and \ac{LiFE-CD}.

\subsubsection{Acyclic Network}
Fig.~\ref{fig:convergence2} presents the corresponding results 
for the acyclic network shown in Fig.~\ref{fig:J1} with fixed link 
failure probabilities $p_{12} = \qty{5}{\percent}$, $p_{23} = \qty{20}{\percent}$, $p_{24} = \qty{20}{\percent}$, while $p_{45}$ is varied over $[0,1)$. 
In contrast to the cyclic case, the Monte Carlo simulation 
aligns with \ac{LiFE-CD} across the entire range of $p_{45}$, confirming that \ac{LiFE-CD} computes the exact expected convergence time for acyclic networks. The bound in~\cite{bernoulli} again exhibits a substantial gap, which overestimates the expected consensus time.

\subsubsection{Variance Analysis}
The simulation results in Fig. \ref{fig:convergence} exhibit notable variance of the expected convergence time, which is inherent to cyclic networks: the presence of cycles introduces  multiple competing propagation paths, such that individual link failure realizations lead to highly variable convergence scenarios. In contrast, this variance is substantially reduced in Fig.~\ref{fig:convergence2} compared to Fig.~\ref{fig:convergence}, since acyclic topologies admit only a single propagation path per source-destination pair, constraining the set of possible convergence scenarios. Unlike the Monte Carlo simulation, \ac{LiFE-CD} computes the probability distribution analytically and is therefore unaffected by such variances.

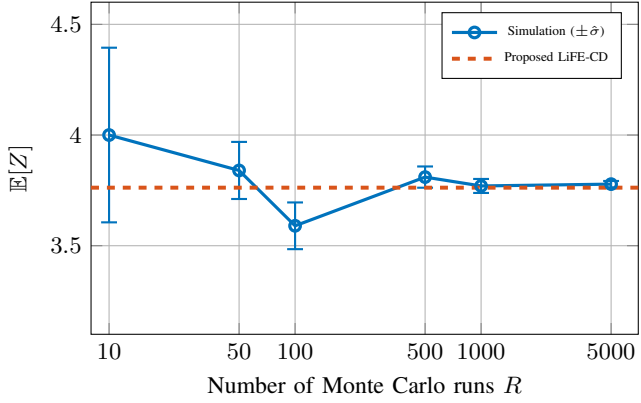
\begin{figure}[t]
\centering
\begin{tikzpicture}
\begin{semilogxaxis}[
    width=\columnwidth,
    height=6cm,
    xlabel={Number of Monte Carlo runs $R$},
    ylabel={$\mathbb{E}[Z]$},
    xmin=8, xmax=7000,
    ymin=3.1, ymax=4.6,
    xtick={10, 50, 100, 500, 1000, 5000},
    xticklabels={$10$, $50$, $100$, $500$, $1000$, $5000$},
    grid=both,
    grid style={line width=0.3pt, draw=gray!30},
    major grid style={line width=0.4pt, draw=gray!60},
    legend pos=north east,
    legend style={
        font=\tiny,
        draw=black,
        fill=white,
        line width=0.4pt,
        inner sep=3pt,
        row sep=1pt,
    },
    tick label style={font=\small},
    label style={font=\small},
    clip=true,
]

\addplot[
    color={rgb,1:red,0.00;green,0.45;blue,0.74},
    mark=o,
    mark size=2pt,
    line width=1.2pt,
    error bars/.cd,
        y dir=both,
        y explicit,
        error bar style={line width=0.8pt, color={rgb,1:red,0.00;green,0.45;blue,0.74}},
        error mark options={
            rotate=90,
            mark size=3pt,
            line width=0.8pt,
        },
] coordinates {
    (10,    4.0000) +- (0, 0.3944)
    (50,    3.8400) +- (0, 0.1289)
    (100,   3.5900) +- (0, 0.1055)
    (500,   3.8100) +- (0, 0.0480)
    (1000,  3.7700) +- (0, 0.0316)
    (5000,  3.7782) +- (0, 0.0143)
};
\addlegendentry{Simulation ($\pm\hat{\sigma}$)}

\addplot[
    color={rgb,1:red,0.85;green,0.33;blue,0.10},
    line width=1.5pt,
    dashed,
    domain=8:7000,
    samples=2,
] {3.76223};
\addlegendentry{Proposed \ac{LiFE-CD}}

\end{semilogxaxis}
\end{tikzpicture}
\caption{Expected convergence time $\mathbb{E}[Z]$ as a function of the
number of Monte Carlo runs~$R$ for the acyclic network in Fig. \ref{fig:J1}. Proposed \ac{LiFE-CD} (red) compared to Monte Carlo simulation (blue) with  error bars indicate one standard deviation~$\hat{\sigma}$.}
\label{fig:averaging}
\end{figure}

To illustrate the number of Monte Carlo runs required to obtain an accurate estimate of the expected convergence time, Fig.~\ref{fig:averaging} shows the sample mean of $\mathbb{E}[Z]$ as a function of the number of runs~$R$ for the acyclic network in Fig. \ref{fig:J1} with link failure probabilities $p_{12} = \qty{5}{\percent}$, $p_{23} = \qty{20}{\percent}$, $p_{24} = \qty{20}{\percent}$ and $p_{45} = \qty{30}{\percent}$. The error bars indicate one standard deviation~$\hat{\sigma}$ for the Monte Carlo simulation. The results reveal that as $R$ increases, the sample mean of the simulation (blue) converges toward the analytical \ac{LiFE-CD} value (dashed red, $\mathbb{E}[Z] = 3.762$) with decreasing standard deviation values, and the confidence interval narrows at a rate of $\mathcal{O}(1/\sqrt{R})$. Consequently, achieving a confidence interval below $\epsilon$ requires $R = \mathcal{O}(\epsilon^{-2})$ simulation runs \cite{O_N_1, O_N_2}, whereas \ac{LiFE-CD} computes $\mathbb{E}[Z]$ analytically in a single run. Unlike Monte Carlo simulation, \ac{LiFE-CD} improves reproducibility by avoiding stochastic effects as occur in simulations.

%% file: performance_2.tex
\subsubsection{Proposed \ac{LiFE-CD} Algorithm}
Since the consensus time $Z$ can in principle take arbitrarily 
large values, the \ac{PMF} and \ac{CDF} arrays are truncated at 
a finite length $N_{\max}$, chosen such that $F_Z(N_{\max}) \approx 1$, 
i.e., the probability of consensus requiring more than $N_{\max}$ 
rounds is negligible. The algorithm proceeds in four phases:

\textit{Phase~1 (Graph Preprocessing):} Dijkstra's algorithm constructs 
$\mathcal{P}_{i^*,\text{SPT}}$ in $\mathcal{O}(n^2)$ for dense graphs or 
$\mathcal{O}(|\mathcal{E}| + n\log n)$ for sparse graphs~\cite{Dijkstra}, where $n = |\mathcal{N}|$ denotes the number of nodes. Identifying 
critical paths via pairwise subpath comparisons requires at most
$\mathcal{O}(n^2)$, giving a total cost of $\mathcal{O}(n^2)$.

\textit{Phase~2 (Geometric Distribution Initialization):} The dominant 
cost of this phase is the computation of the geometric \ac{CDF}, where 
one geometric \ac{CDF} of length $N_{\max}$ is computed per leaf node in 
$\mathcal{O}(N_{\max})$, yielding $\mathcal{O}(n \cdot N_{\max})$ in total.

\textit{Phase~3 (Iterative Network Reduction):} The loop runs at most 
$n-1$ iterations. In each iteration, convolution for unicast transmission 
mode costs $\mathcal{O}(N_{\max}^2)$, where the result is truncated 
to length $N_{\max}$ after each step to maintain fixed array size. Pointwise products of \ac{CDF}s 
for broadcast transmission costs $\mathcal{O}(n \cdot N_{\max})$. Thus, 
the dominant cost per iteration is
\begin{equation}
    \mathcal{O}\bigl(N_{\max} \cdot \max(n,\, N_{\max})\bigr).
\end{equation}
Since the number of iterations is at most $n-1$, the total cost is given by
\begin{equation}
    \mathcal{O}\bigl(n \cdot N_{\max} \cdot \max(n,\, N_{\max})\bigr).
    \label{eq:O_phase3}
\end{equation}

\textit{Phase~4 (Final Composition):} One final broadcast or unicast 
step contributes $\mathcal{O}(n \cdot N_{\max})$ or 
$\mathcal{O}(N_{\max}^2)$, respectively. As Phase~3 dominates, 
the total complexity of \ac{LiFE-CD} is given by~\eqref{eq:O_phase3}.

\subsubsection{Monte Carlo Simulation}
The communication graph is represented by an adjacency 
matrix, such that in each communication round, each of 
the $n$ nodes determines its active neighbors by accessing 
the corresponding row of the adjacency matrix, resulting 
in a per-round complexity of $\mathcal{O}(n^2)$. Since 
the expected number of rounds is bounded by 
$\frac{n}{1-p_{\max}}$ using~\eqref{eq:golfar_bound} 
with $D \leq n$, and denoting by $R$ the number of Monte 
Carlo runs required to obtain a statistically reliable 
estimate of the convergence time distribution, the 
expected simulation complexity is given by
\begin{equation}
    \mathcal{O}\!\left(\frac{n^{3} \cdot R}{1-p_{\max}}\right).
    \label{eq:O_MC_combined}
\end{equation}

\subsubsection{Complexity Comparison}
To compare the computational complexity of \ac{LiFE-CD} and 
Monte Carlo simulation, the truncation length $N_{\max}$ 
used in \ac{LiFE-CD} must be bounded. By Markov's inequality, 
$\Pr(Z > N_{\max}) \leq \mathbb{E}[Z]/N_{\max}$, and 
substituting the upper bound~\eqref{eq:golfar_bound} with 
$D \leq n$ yields $\Pr(Z > N_{\max}) \leq \frac{n}{(1-p_{\max})\,N_{\max}}$. 
To ensure a truncation error of at most $\epsilon > 0$, 
it suffices to choose $N_{\max} \geq \frac{n}{(1-p_{\max})\,\epsilon}$, 
giving
\begin{equation}
    N_{\max} \in \mathcal{O}\!\left(\frac{n}{1-p_{\max}}\right).
    \label{eq:N_max_scaling}
\end{equation}
Substituting~\eqref{eq:N_max_scaling} into~\eqref{eq:O_phase3} and 
using $\max(n,\,N_{\max}) \leq n + N_{\max}$, the \ac{LiFE-CD} 
complexity becomes
\begin{align}
    \mathcal{O}\bigl(n \cdot N_{\max} \cdot (n + N_{\max})\bigr)
    &= \mathcal{O}\!\left(\frac{n^3}{1-p_{\max}} + 
       \frac{n^3}{(1-p_{\max})^2}\right) \notag\\
    &= \mathcal{O}\!\left(\frac{n^3}{(1-p_{\max})^2}\right),
    \label{eq:complexity_LiFE}
\end{align}
where the second term dominates since $p_{\max} \in (0,1)$. Since~\eqref{eq:complexity_LiFE} is a worst-case 
bound whereas~\eqref{eq:O_MC_combined} is an expected complexity 
that may be exceeded in practice, this comparison is conservative, 
yet \ac{LiFE-CD} outperforms Monte Carlo simulation whenever
\begin{equation}
    \frac{n^3}{(1-p_{\max})^2} < \frac{n^3 \cdot R}{1-p_{\max}}
    \;\Leftrightarrow\; p_{\max} < 1 - \frac{1}{R},
    \label{eq:speedup_condition}
\end{equation}
with speedup factor $\mathcal{O}(R \cdot (1-p_{\max}))$. This 
condition is satisfied for any statistically meaningful simulation, 
e.g., $p_{\max} < 0.99$ for $R = 100$ and $p_{\max} < 0.999$ for 
$R = 10^3$. As Monte Carlo requires large $R$ to maintain 
statistical reliability, \ac{LiFE-CD} is computationally superior 
in all practically relevant scenarios while providing exact and tightly bounded
distributions rather than statistical estimates. The overall 
comparison is summarized in Table~\ref{tab:complexity_comparison}.

\begin{table}[ht]
\centering
\renewcommand{\arraystretch}{1.2}
\caption{Comparison of key characteristics of Monte Carlo simulation and the proposed \ac{LiFE-CD} algorithm.}
\begin{tabular}{lcc}
\hline
\textbf{Metric} & \textbf{Simulation} & \textbf{LiFE-CD} \\
\hline
Complexity & $\mathcal{O}(\frac{n^3 \cdot R}{1-p_{\max}})$ & $\mathcal{O}(\frac{n^3}{(1-p_{\max})^2})$ \\
Randomness & Yes & No \\
Depends on averaging & Yes & No \\
Runtime depends on $p_{ij}$  & Yes & No \\
Reproducibility & Low & High \\
\hline
\end{tabular}%
\label{tab:complexity_comparison}
\end{table}

%% file: conclusion.tex
This paper proposed an analytical convergence time analysis of the max-consensus protocol in multi-agent systems under Bernoulli-distributed link failures. The key contribution is a novel algorithm, termed \ac{LiFE-CD}, that deterministically 
computes the full probability distribution of convergence time from network topology and individual link failure probabilities. By modeling per-link delays as geometric random variables, the network is iteratively reduced via convolution for unicast and pointwise multiplication for broadcast transmissions.

For acyclic networks, the proposed \ac{LiFE-CD} yields exact results, 
validated by comprehensive Monte Carlo simulations. For cyclic networks, tight upper bounds on the probability distribution are obtained via shortest-path spanning tree construction, which significantly improve consensus time estimation compared to existing bounds. Furthermore, the full probability distribution enables deadline-aware protocol design with specified reliability guarantees. Our complexity analysis demonstrates superiority over Monte Carlo simulation, while eliminating stochastic variability and ensuring reproducibility. All results extend directly to min-consensus by structural equivalence.

Future work will investigate extensions to time-varying network topologies, correlated link failures, and heterogeneous agent update rates, as well as the application of \ac{LiFE-CD} to deadline-constrained protocol optimization in cooperative \ac{UAV} swarms or robots, and industrial wireless sensor networks.